# Pressure-induced structure transformation and collapse of ferromagnetism in van der Waals insulator

*M. Míšek[1], U. Dutta[1], P. Král[2], D. Hovančík[2], J. Kaštil[1], K. Pokhrel[2], S. Ray[2], J. Valenta[1,2] J. Prchal[2], J. Kamarád[1], F. Borodavka[3], V. Eigner[1], M. Dušek[1], V. Holý[2], K. Carva[2], S. Kamba[3], V. Sechovský[2] and J. Pospíšil[2]*

[1]Institute of Physics of the Czech Academy of Sciences, Cukrovarnická 10/112, 162 00 Prague 6, Czech Republic

[2] Charles University, Department of Condensed Matter Physics, Ke Karlovu 5, 121 16 Prague 2, Czech Republic

[3]Institute of Physics of the Czech Academy of Sciences, Na Slovance 2, 182 00 Prague 8, Czech Republic

**Abstract**

Ferromagnetism in van der Waals insulators like $CrBr_3$ is highly sensitive to structural modifications. We explore the pressure-driven structural and magnetic transformations of the van der Waals magnet $CrBr_3$, establishing it as a model platform for phenomena emerging in layered vdW systems. Single-crystal X-ray diffraction revealed intrinsic trimorphism with two known phases (monoclinic and rhombohedral) and a so-far unreported trigonal phase. The paracrystal model well captures the coexistence of the rhombohedral and trigonal phases at ambient conditions. Increasing pressure drives the growth of the AA-stacked trigonal phase at the expense of the rhombohedral phase, which becomes undetectable above 8.4 GPa. Magnetization measurements provided the first direct evidence of a collapse of ferromagnetism above 5.8 GPa. This behavior is attributed to the increasing number of AFM-coupled Cr moments in AA stackings. *Ab initio* DFT calculations of electronic structure and atomistic simulations of finite-temperature magnetism corroborate the scenario.

## INTRODUCTION

The very weak van der Waals (vdW) interlayer bonding in vdW materials allows for the exfoliation of a crystal down to a monolayer [1]. Demonstrating intrinsic ferromagnetism (FM) in a monolayer [2] has opened new gateways for research and applications of 2D magnetic materials [3-6]. Applying mechanical stimuli, especially external strain or pressure, can unveil the delicate balance of the inter- and intra-layer exchange interactions. The Cr trihalides $CrX_3$ ($X$ = Cl, Br, I) are unique in this context. [7-15].

The exclusive feature of all three $CrX_3$ compounds is their structural dimorphism, which naturally allows for two different layer stackings [16, 17]. At high temperatures, they adopt the monoclinic $AlCl_3$ structure type ($C2/m$ space group) that transforms upon cooling to the rhombohedral $BiI_3$-type structure ($R\overline{3}$) by a first-order phase transition [18, 19]. The common structural motif is the honeycomb net of Cr cations in edge-sharing octahedral coordination. The layer stacking sequence in the $BiI_3$ type is strictly ABC, whereas the layer stacking in the $AlCl_3$ type is approximately ABC. In the former case, the subsequent layers are shifted along one of the Cr-Cr bonds, while in the latter case, the layers are moved perpendicular to this direction [16].

In the $CrI_3$ rhombohedral phase, monoclinic stacking faults were detected [20, 21], accompanied by an antiferromagnetic (AFM) contribution in the magnetization isotherms [22, 23].

Hydrostatic pressure was reported to suppress ferromagnetism in $CrI_3$, and the formation of an unknown high-pressure AFM phase was suggested [8].

$CrBr_3$ is a ferromagnet with Curie temperature $T_C$ = 33 K [24]. The structural transition occurs upon cooling at 420 K [16]. In contrast to $CrI_3$, no AFM contribution was detected in magnetization isotherms measured in ambient pressure [18, 24]. However, the sensitivity of ferromagnetism in $CrBr_3$ to adjacent-layer misfits was demonstrated by magnetoconductance measurements, which revealed multiple AFM phases [25]. Magnetic order in $CrBr_3$ persists to monolayer thickness with a $T_C$ found to be 22 K [26] or 27 K [27, 28].

Several previous publications on $CrBr_3$ have dealt with the influence of external pressure on ferromagnetism and crystal structure [9-13]. First, $T_C$ was reported to decrease linearly with pressure up to 1 GPa with $\partial T_C/\partial P$ = - 2 K/GPa [9]. Lis et al. [12] observed that $T_C$ decreased with increasing pressure up to 3 GPa. From a linear extrapolation of the 0 to 3 GPa data to higher pressures, they estimated ~ 8.4 GPa as the critical pressure for the loss of ferromagnetism. They also suggested a gradual isostructural phase transition in the pressure range of 2.5–7 GPa based on powder X-ray diffraction data. A pressure-induced structural transition at 9.5 GPa was inferred from changes in the Raman spectra [13]. The theoretical study of the influence of biaxial strain on atomic and electronic structure predicted that compressive strain can induce a transition from ferromagnetic (FM) to AFM state in a $CrBr_3$ monolayer [29].

The dominant nearest-neighbor exchange interaction originates in the superexchange mechanism between 2 Cr atoms via Br. Since the angle of the Cr-Br-Cr bond is close to 90˚ (as is typically the case for $CrX_3$ vdW magnets), this interaction should favor FM ordering according to the Goodenough-Kanamori-Anderson rules [30]. DFT studies predict an increase in the nearest-neighbor exchange in $CrBr_3$ with pressure, which is connected to the fact that the calculated Cr-Br-Cr angle decreases from 94˚ towards the ideal value 90˚ [11]. Interlayer interaction is of high importance in layered vdW (quasi-2D) magnets as it significantly affects critical temperatures[3, 31]. It also increases with pressure, probably due to increasing overlap between distant orbitals. Therefore, the Curie temperature is expected to grow. Another study predicted that a different magnetic order would occur at 17 GPa, but data under 10 GPa were not shown. The observed decrease of the Curie temperature with pressure thus represents a long-standing puzzle.

To achieve a comprehensive understanding of the effects of external pressure on the crystal structure and FM of $CrBr_3$, we performed a thorough experimental investigation of magnetization, single-crystal X-ray diffraction (XRSCD), reciprocal-space mapping, and Raman spectroscopy, in conjunction with *ab initio* DFT calculations of electronic structure and atomistic simulations of finite-temperature magnetism. The diffraction measurements in ambient pressure and at room temperature already revealed a new trigonal phase ($P\overline{3}m1$ space group) coexisting with the only known rhombohedral phase ($R\overline{3}$). We then described the complex process of the pressure-induced structural phase transformations from the initially dominant $R\overline{3}$ phase to the final single high-pressure phase $P\overline{3}m1$. The main achievements of our work are a) the presentation of the first direct experimental proof of the pressure-induced suppression of FM facilitated by challenging magnetization measurements at high pressures and b) the explanation of the striking loss of FM ordering of localized Cr magnetic moments by introducing the AFM interactions gradually emerging in the magnetic lattice intertwined with extending the layer sequences of the new trigonal $P\overline{3}m1$ phase with increasing pressure.

## RESULTS AND DISCUSSION

### ***High-pressure magnetization measurement on $CrBr_3$ single crystal***

The thermomagnetic curves and magnetization isotherms measured on single crystals under several external pressures in magnetic fields up to 1 T applied out-of-plane (along the crystallographic *c*-axis) are shown in Fig. 1. It is observed that both $T_C$ and saturated magnetization $M_{sat}$ decrease from the ambient pressure values $T_C = 33$ K and $M_{sat} \sim 2.8\ \mu_B$/f.u. at 2 K in an identical trend. The phase diagram in Fig. 1 shows that $T_C$ decreases moderately with pressure up to 3 GPa, where the $T_C$ vs. $p$ fall rate gradually accelerates, leading to $T_C \approx 4$ K at $p = 5.6$ GPa. The extrapolated critical pressure is $p_c = 5.8$ GPa. The measurement at 6.24 GPa showed no FM signal.

The critical pressure of 5.8 GPa from our experiment is much lower than 8.4 GPa predicted from a linear extrapolation of the 0 to 3 GPa data [12]. The $p_c$ value is somewhat sample-dependent. Two experimental runs with different $p_c$ are presented in Fig. S1a [32]. The FM state with unaffected $T_C$ and $M_{sat}$ at 2 K recovers after decompression from 10 GPa (see Fig. S1b [32]). The gradual suppression of FM in $CrBr_3$ and $CrI_3$ [8], revealed by our preliminary magnetization measurements (see Fig. S2 [32]), is striking. Both materials are insulators with well-localized Cr 3*d* electrons bearing magnetic moments. $T_C$ of such ferromagnets is expected to increase weakly with increasing pressure [33] while the magnetic moments remain intact. This expectation is based on the presumption of a standard elastic response and the persisting symmetry of the crystal lattice and the hierarchy of exchange interactions. However, $T_C$ and spontaneous magnetization in $CrI_3$ and $CrBr_3$ are found to decrease rapidly with increasing pressure and to collapse of ferromagnetism at several GPa, in contrast to expectations. The spontaneous magnetization and $T_C$ decrease rapidly with pressure, a common characteristic of metallic ferromagnets with delocalized/itinerant *d*-electrons, which is certainly not the case for the insulators $CrI_3$ or $CrBr_3$.

Kozlenko et al. observed a negative thermal expansion at temperatures below $T_C$ [34], which would imply a negative $\partial T_C/\partial P$ in hydrostatic pressure when considering the Ehrenfest relation:

$$\frac{\partial T_C}{\partial P} = \frac{\Delta\kappa_{T_C}}{3\Delta\alpha_{T_C}} \quad (1),$$

where the terms $\Delta\kappa_{T_C}$ and $\Delta\alpha_{T_C}$ represent the discontinuity in volume compressibility and thermal expansion below $T_C$, respectively. However, this approach treats magnetovolume effects in materials with almost isotropic elastic properties and FM interactions. These conditions certainly do not apply to vdW ferromagnets; the simple thermodynamic approach cannot correctly explain the pressure-induced suppression of ferromagnetism in $CrBr_3$.

Some experimental findings for $CrI_3$ indicate that hydrostatic pressure inhomogeneously modifies the stacking order [35, 36], potentially suppressing ferromagnetism by pressure-induced inclusions of layer stackings with AFM elements. Similar indications have been obtained by investigations of $CrBr_3$ misfit bilayers [25]. We discuss these mechanisms together with the results of the X-ray diffraction investigation of the crystal structure and the theoretical calculations.

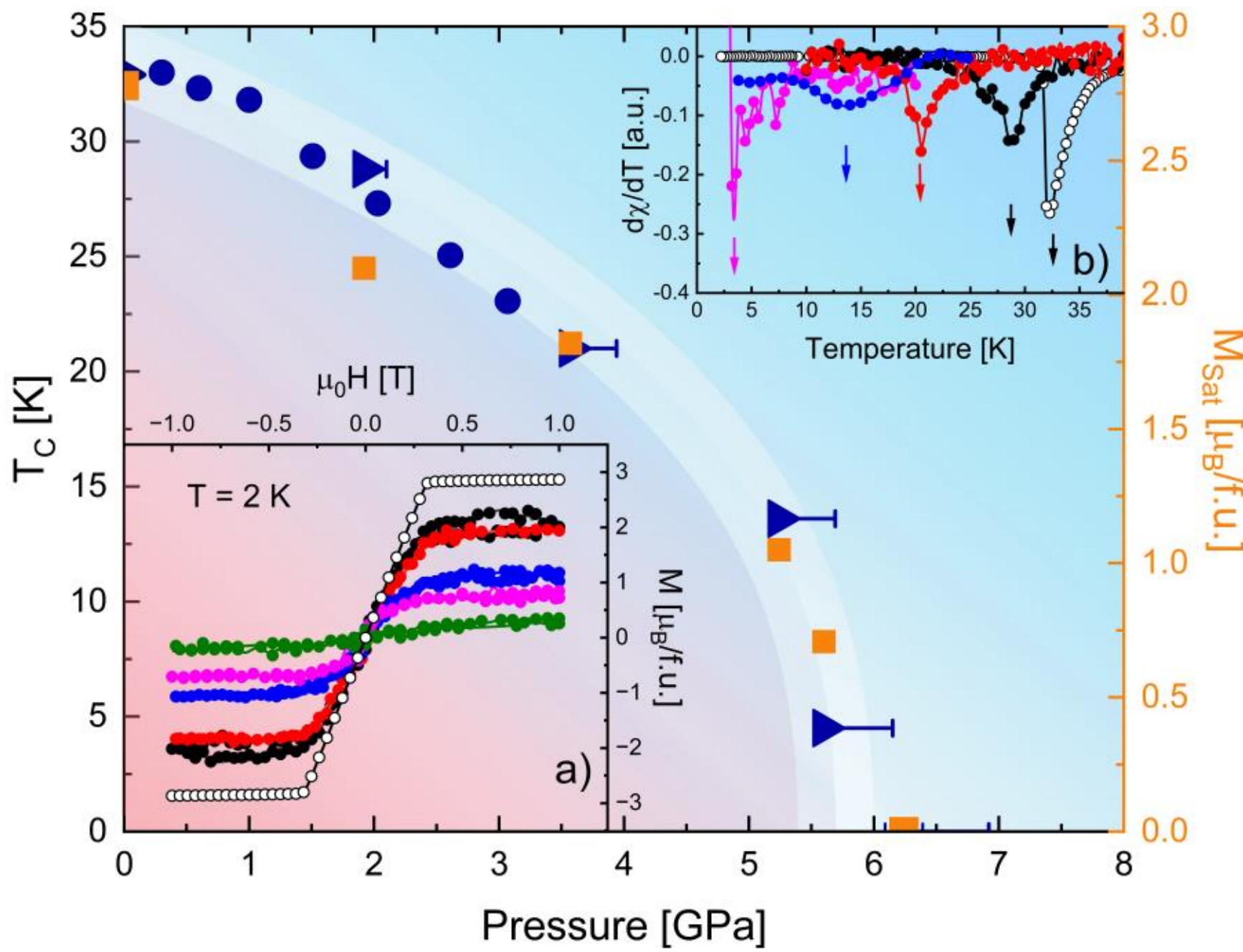

**Fig. 1 | The magnetic *p-T* phase diagram of $CrBr_3$.** The phase diagram shows pressure dependences of the Curie temperature and saturated magnetization. The evolution of the Curie temperature was constructed based on low-pressure AC susceptibility measurement ● and DC magnetization measurement in a DAC cell ▶ in a magnetic field of 0.01 T. The horizontal right-side error bars show the overpressure inside the DAC when setting the experiment point. Saturated magnetization $M_{sat}$ ■ was read as the 1 T magnetization value. The insets show source data from DAC with the joined legend scheme **O** 0 GPa, ● 1.92 GPa, ● 3.57 ● 5.24 GPa, ● 5.60 GPa, and ● 6.24 GPa. **a** Magnetization loops for the detection of the saturated magnetization $M_{sat}$. **b** M(T) dependences measured at a low field of 0.01 T applied out-of-plane (along the crystallographic *c*-axis).

### *Crystal structure analysis*

<u>*Ambient pressure structure*</u>

The experiment at ambient pressure confirmed the expected $BiI_3$-type rhombohedral $R\overline{3}$ structure reported in the literature [16, 37]. During refinement of the ambient pressure structure ($R\overline{3}$), significant scattering density was detected in the ideally vacant *3a* site centers [34] (the Cr2 sites) of the Cr honeycombs. Hence, extra chromium ions have been introduced into the refined crystal structure at the *3a* site to specify the crystal structure averaged over the layers. The occupancies of the Cr1 and Cr2 sites were refined with the sum constrained to 1, leading to 0.878(4) and 0.244(8), respectively (Figs. 2a and 2b). The similar extra scattering density in the theoretically ideally vacant centers at the *3a* site was also found in vanadium trihalides $VI_3$ and $VBr_3$ [38, 39] in the rhombohedral $R\overline{3}$ structure. In both cases, this result was attributed to stacking faults without deeper analysis.

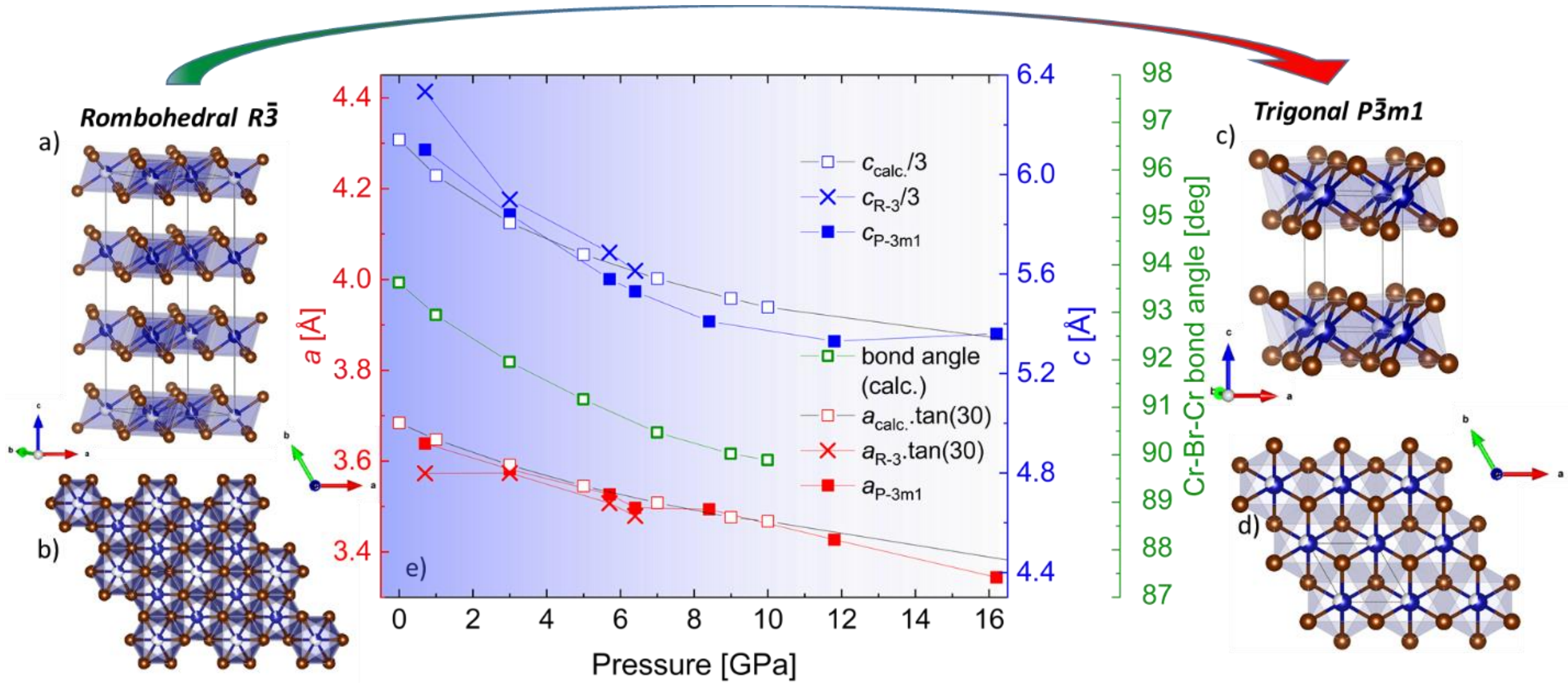

**Fig. 2 | Results of the X-ray diffraction study-$CrBr_3$ crystal structure**. The black-line polyhedron represents the unit cell. The blue sections on the Cr sites graphically demonstrate the occupancy, and the brown balls represent the Br sites. **a** The unit cell of the rhombohedral $R\overline{3}$ phase is three times longer along the *c*-axis than **c** the trigonal $P\overline{3}m1$. **d** The 2/3 occupied Cr sites are arranged in AA stacking in $P\overline{3}m1$ **b** while the ABC sequence of adjacent layers is present within the $R\overline{3}$ structure. **e** Lattice parameters of $CrBr_3$ as a function of pressure at room temperature. For results in $P\overline{3}m1$, we have used $R\overline{3}$ convention of the unit cell. The calculated lattice parameters *a* and *c,* and the Cr-Br-Cr angle as a function of pressure, are also included.

<u>*Structure evolution with pressure*</u>

The pressure causes a gradual breakdown of the $R\overline{3}$ structure that seems to vanish at a pressure between the two consecutive measurements at 6.4 GPa, at which the 1/3 satellite reflections are still visible (Figs. 3a-b), and 8.4 GPa (Fig. 3c), where they are extinct within the detection limit of the method. X-ray data affected by the absorption of the pressure cell, however, does not allow for a precise quantitative analysis of the $R\overline{3}$ / $P\overline{3}m1$ content ratio. When the data are interpreted as a stacking violation, the situation is even more complex; we can determine reliably only general trends. The key finding of this experiment is that the high-pressure induced trigonal phase ($P\overline{3}m1$ space group) remains stable after decompression to ambient pressure from the last measured point at 21.5 GPa (see Fig. 3d and Tab. S4). This result suggests that the $P\overline{3}m1$ phase has been stabilized at the highest pressure and preserved during decompression to ambient pressure, where it remains metastable.

The high-pressure phase is trigonal with $P\overline{3}m1$ symmetry and a single Cr site of 2/3 occupancy at *1a* position. Full crystallographic solutions of $R\overline{3}$ and $P\overline{3}m1$ phases are summarized in Tables S1-S3 [32], respectively. Fig. 2e graphically presents the pressure dependence of the lattice parameters of $R\overline{3}$ and $P\overline{3}m1$ phases. Both lattice parameters shorten with increasing pressure and follow a slightly convex dependence. The shrinking of the lattice parameter *c* saturates above the $p \sim 11.8$ GPa, which has been attributed to the 3D character of the $P\overline{3}m1$ phase at higher pressures [40]. The corresponding experimental data are listed in Table S4 [32].

We note that all single crystals measured at ambient pressure always exhibited a finite fraction of the $P\overline{3}m1$ polymorph, as evidenced by the characteristic fading of the "1/3" reflections.

The relative abundance of $P\overline{3}m1$ phase appeared to be sensitive to, and likely unintentionally modified by, subsequent sample manipulation during mounting inside the pressure cell.

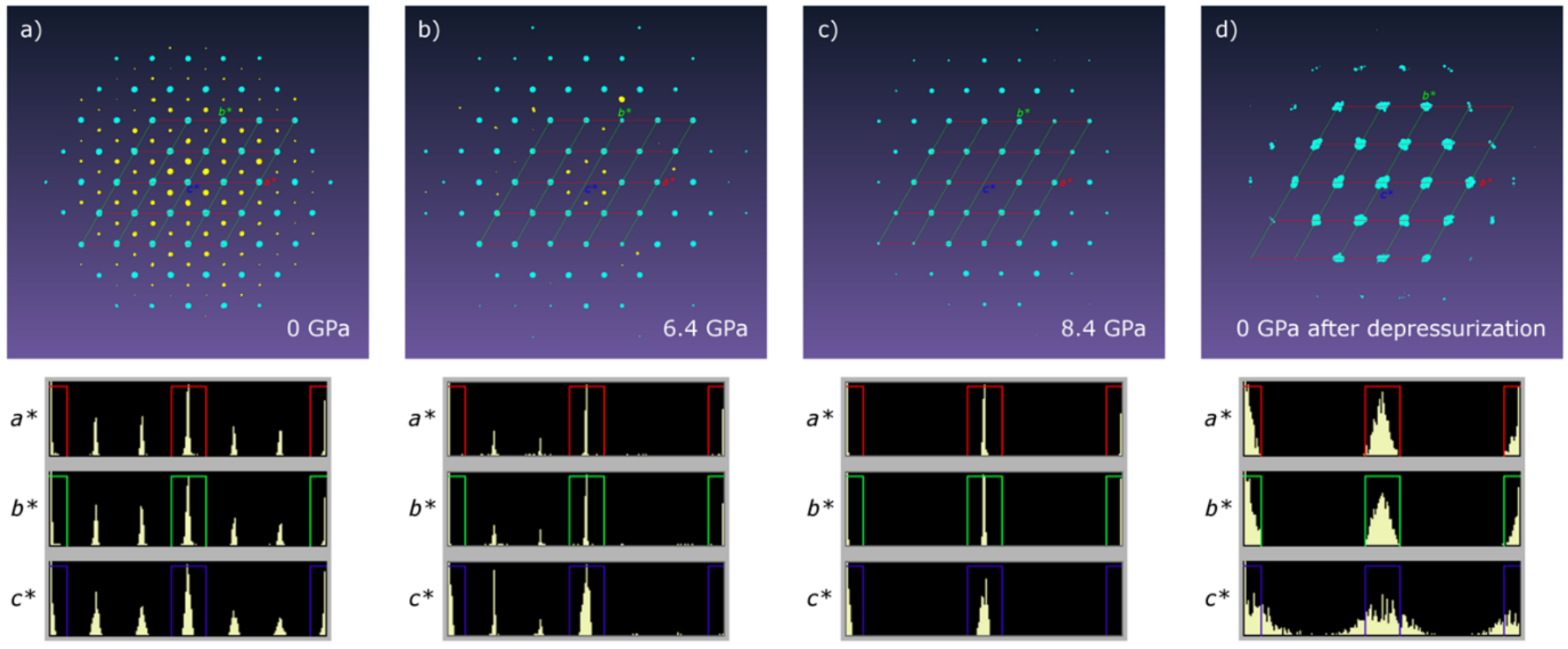

**Fig. 3 | Reciprocal space reconstructions of CrBr3 single crystal diffraction data** viewed in the direction of $c*$ axis at **a** ambient pressure, **b** lower, **c** higher than the transition pressure, and **d** after decompression from 21.5 GPa. The diffractions belonging to both $R\overline{3}$ and $P\overline{3}m1$ phases are depicted in cyan, diffractions belonging only to $R\overline{3}$ phase in yellow. The lower trio of figures represents distribution histograms for $a^* = b^*$ and $c^*$ vector projections.

The two phases are indistinguishable by powder X-ray diffraction because the most intense corresponding reflections practically overlap. That is why the $P\overline{3}m1$ phase has remained unresolved in all previous structure studies [11, 12, 40] (see the comparison of the X-ray diffraction patterns in Fig. S3 [32].

An alternative interpretation of the observed X-ray patterns describing the high-pressure induced hexagonal structure with 2/3 occupancy of Cr sites (Figs. 2c and 2d) may be explained in terms of the gradual extinction of the characteristic "1/3" peaks pointing to the destabilization of the $R\overline{3}$ rhombohedral structure by lattice stresses induced by external pressure. In this structural model, the basal plane motif remains unaltered. Still, the placement of the adjacent layer is randomized along the $c$-axis, leading to seemingly 2/3 occupancy when integrated through the entire bulk of the crystal. Interestingly, the positions of lattice points in both phases $R\overline{3}$ and $P\overline{3}m1$ almost coincide; if we neglect local lattice distortions around the Cr sites, which result in slightly different Br positions around occupied and unoccupied Cr sites, the lattices are identical. They differ only in the Cr occupancies. The $R\overline{3}$ phase exhibits a periodic sequence of Cr occupations (0.878 in four 6c Wyckoff positions and 0.244 in two 3a positions in one unit cell); this leads to effective ABC stacking; on the other hand, the Cr occupancies in the $P\overline{3}m1$ are identical, which results in the AAA stacking. The pressure-induced $R\overline{3} \rightarrow P\overline{3}m1$ phase transformation consists only of the randomization of Cr occupancies and consequent small local changes in Br positions. An alternative scenario involves randomizing the vertical stacking of the entire A, B, C basal planes. This process, in effect, leads to the average AAA stacking again. From the current X-ray diffraction data, we are unable to distinguish between these two scenarios.

*Analysis of the layers order*

A thorough experimental study of the stacking of the $CrBr_3$ layers in as-grown single crystals at ambient conditions was performed using X-ray diffraction curves (symmetric 2θ/ω scans) and measuring high-resolution reciprocal-space maps. An as-grown $CrBr_3$ single crystal of size ~10 mm$^2$ was chosen for the experiment; inside a glovebox, under Ar environment, the single crystal was placed on a Si non-diffracting single-crystalline plate and covered by a blue exfoliation tape to protect the sample from the environment.

The calculated model of the expected reciprocal lattice for both crystallographic phases of $CrBr_3$ is displayed in Fig. 4a. Both lattices are very well distinguishable; the $R\overline{3}$ phase exhibits additional diffraction maxima outside the symmetric 000L rod.

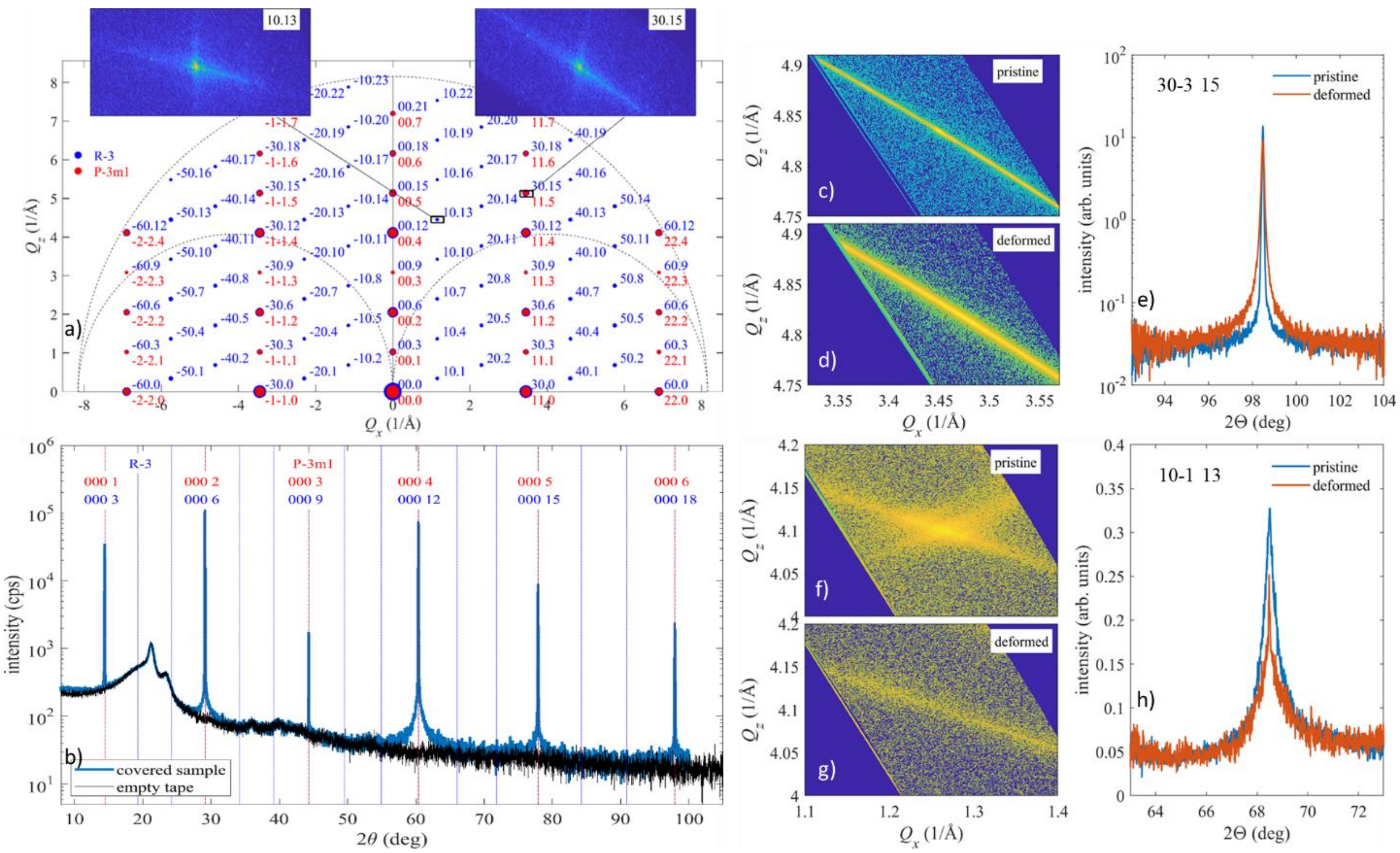

**Fig. 4 | Results of the X-ray diffraction studies of the $CrBr_3$ layer stacking. a** Calculated model of the reciprocal lattice of the $CrBr_3$ rhombohedral phases $R\overline{3}$ (space group 148) blue reciprocal-lattice points and $P\overline{3}m1$ (space group 150 – red points). The size of the circular symbols roughly corresponds to the relative intensities of corresponding reflections, and the dotted arcs denote the boundaries of the reciprocal space region accessible in the used coplanar reflection scattering geometry. The inset displays the reciprocal-space maps measured around the asymmetric maxima $10\overline{1}.13$ and $30\overline{3}.15$. **b** Measured symmetric 2θ/ω scans of the $CrBr_3$ single crystal covered by the exfoliation tape (blue) and of the empty tape (black). The vertical red dotted lines represent the expected positions of the diffraction maxima of the $P\overline{3}m1$ and $R\overline{3}$ structures, along with their reflection indexes; the blue dotted lines denote the positions of the maxima $000.(3n+1)$ and $000.(3n+2)$ forbidden in both structures. Reciprocal space maps measured around the reciprocal lattice point $30\overline{3}.15$ and $10\overline{1}.13$ before **c,f,** and after **d,g** deformation, as well as intensities integrated along the ω-circles **e,h**, respectively.

Fig. 4b shows the 2θ/ω diffraction curve measured in the symmetric reflection geometry, i.e., crossing the symmetric reciprocal-lattice points $000.L$. The symmetric scan presented here cannot distinguish the structures under consideration since the diffraction maxima $000.(3n+1)$ and $000.(3n+2)$ are forbidden in both cases. The width of the diffraction peaks is comparable to the device resolution, indicating the high quality of the individual mosaic crystallites composing the sample.

The insets in Fig. 4a show the reciprocal-space maps measured around the reciprocal lattice points $30\bar{3}.15$ (for both $R\bar{3}$ and $P\bar{3}m1$ phases) and $10\bar{1}.13$ (only $R\bar{3}$ phase). The inclined streaks in the maps (perpendicular to the corresponding reciprocal-lattice vectors) indicate crystal mosaicity (the misorientation of the individual crystallites lies below 1 degree). The vertical streaks are probably caused by the crystal surface (crystal truncation rods). From the existence of the $10\bar{1}.13$ maximum, we conclude that the $R\bar{3}$ phase is present in the investigated sample; however, from this data, the presence of the other phase $P\bar{3}m1$ cannot be excluded.

Therefore, to investigate the structural changes due to pressure, we measured X-ray diffraction reciprocal space maps around the reciprocal lattice points $10\bar{1}.13$ and $30\bar{3}.15$ before and after deformation by an external force. In Figs. 4c,d, and 4f,g, respectively. We show the reciprocal-space maps before and after deformation, as well as the intensities from the map integrated along the Debye circles (i.e., keeping the scattering angle 2θ constant) (Figs 4e and 4h). The $30\bar{3}.15$ maximum was slightly broadened by deformation, but it retained its Lorentzian-like shape. On the contrary, the $10\bar{1}.13$ diffraction peak changed its form substantially; after deformation, it exhibits a central sharp maximum accompanied by a broader lower component.

This behavior can be explained by the paracrystal model [41-43], which describes the sample as a random sequence of blocks of $R\bar{3}$ and $P\bar{3}m1$ phases. Fig. S4 [32] shows the results of simulations in which we used various mean lengths of the blocks and assumed that the block lengths are distributed according to a Poisson distribution. Indeed, the shape of the 30-3 15 peak is almost independent of the block lengths, while the width of the broad component of the 10-1 13 maximum is inversely proportional to the mean block length. From the width of the broad component in Fig. 4h [32], we can estimate the mean width of the blocks as $(10 \pm 2)$nm.

A major limitation of our high-pressure XRD data is the low resolution, which prevents a direct determination of the coherent domain size. The intensity of the 1/3 reflections depends on both the phase volume fraction and the occupancy difference between the 6c and 3a Wyckoff positions. The phase volume ratio is determined from the relative intensities of the 1/3 maxima, assuming the presence of only "pure" phases, i.e., $R\bar{3}$ (occupancies 0.878 and 0.244) and $P\bar{3}m1$ (occupancies 2/3 and 2/3). This represents an extreme case. In a real crystal, however, the $R\bar{3}$ phase may have a somewhat lower occupancy ratio in some regions. The actual volume of the $R\bar{3}$ phase is then larger than the one calculated, meaning that the $R\bar{3}$ volume fraction listed in Table 4S is a lower bound.

***Raman spectroscopy***

The pressure evolution of the phonon excitations in $CrBr_3$ at room temperature, as revealed by Raman spectroscopy, is shown as a map in Fig. 5. The source data are displayed in Fig. S5 [32]. At 0.14 GPa, we observed four characteristic Raman peaks: at 105.7 cm$^{-1}$ (Ag), 140.7 cm$^{-1}$ (Eg), 149.9 cm$^{-1}$ (Eg), and 181.8 cm$^{-1}$ (A2g). These phonons are qualitatively consistent with previous publications [12, 13, 40, 44].

Two principal effects are observed in the Raman spectra. First, the phonon intensities gradually vanish. We attribute this decrease to the contrast in dielectric constants between the $R\overline{3}$ and $P\overline{3}m1$ phases, combined with the progressive metallization of $CrBr_3$ at elevated pressures[13].

The second effect is the pressure dependence of the phonon frequencies. Previous Raman studies attributed all observed phonon excitations to $CrBr_3$, assumed to crystallize solely in the $R\overline{3}$ phase. However, our extensive single-crystal X-ray diffraction analysis of dozens of specimens has never yielded a purely $R\overline{3}$ crystal; a fraction of the $P\overline{3}m1$ phase was always present. Treating the Raman spectrum as an additive superposition of both phases, we can identify two groups of phonon modes (Fig. 5) distinguished by their different pressure-dependent slopes.

All modes $Ph_{6\text{-}8}$ existing above 10 GPa, marked in red lines, are hard to pressure with a slope of ~1.7 cm$^{-1}$/GPa. On the other hand, a mixture of both hard and more responsive soft modes is present below 10 GPa, as indicated by the blue and green lines with slopes ~2.9 ($Ph_4$) and ~4.2 cm$^{-1}$/GPa ($Ph_2$ and $Ph_5$), respectively. For this reason, we assign all hard (red) modes to $P\overline{3}m1$ phase being considered having less compressible 3D character, while the soft (green + blue) modes to 2D $R\overline{3}$ phase[40]. As the $P\overline{3}m1$ phase exists at high pressures; only hard modes are detected at these conditions. While both types of modes appear at lower pressures because of $R\overline{3}$ and $P\overline{3}m1$ mutual coexistence. All soft phonon modes transform to hard modes around 10 GPa, as best documented by the couples $Ph_5$ - $Ph_8$ and $Ph_4$ - $Ph_7$; the intensity of $Ph_1$ is hardly detectable above $p_c$. While the change in trend appears at pressure ~10 GPa in our case; 15 GPa was reported in[40]. We can only speculate that the initial structural state of samples may influence their response at higher pressures.

When both trigonal space groups are very similar in terms of symmetry, no characteristic splitting, merging, or emergence of new phonon branches is expected, as typically observed at structural transitions with symmetry breaking[45].

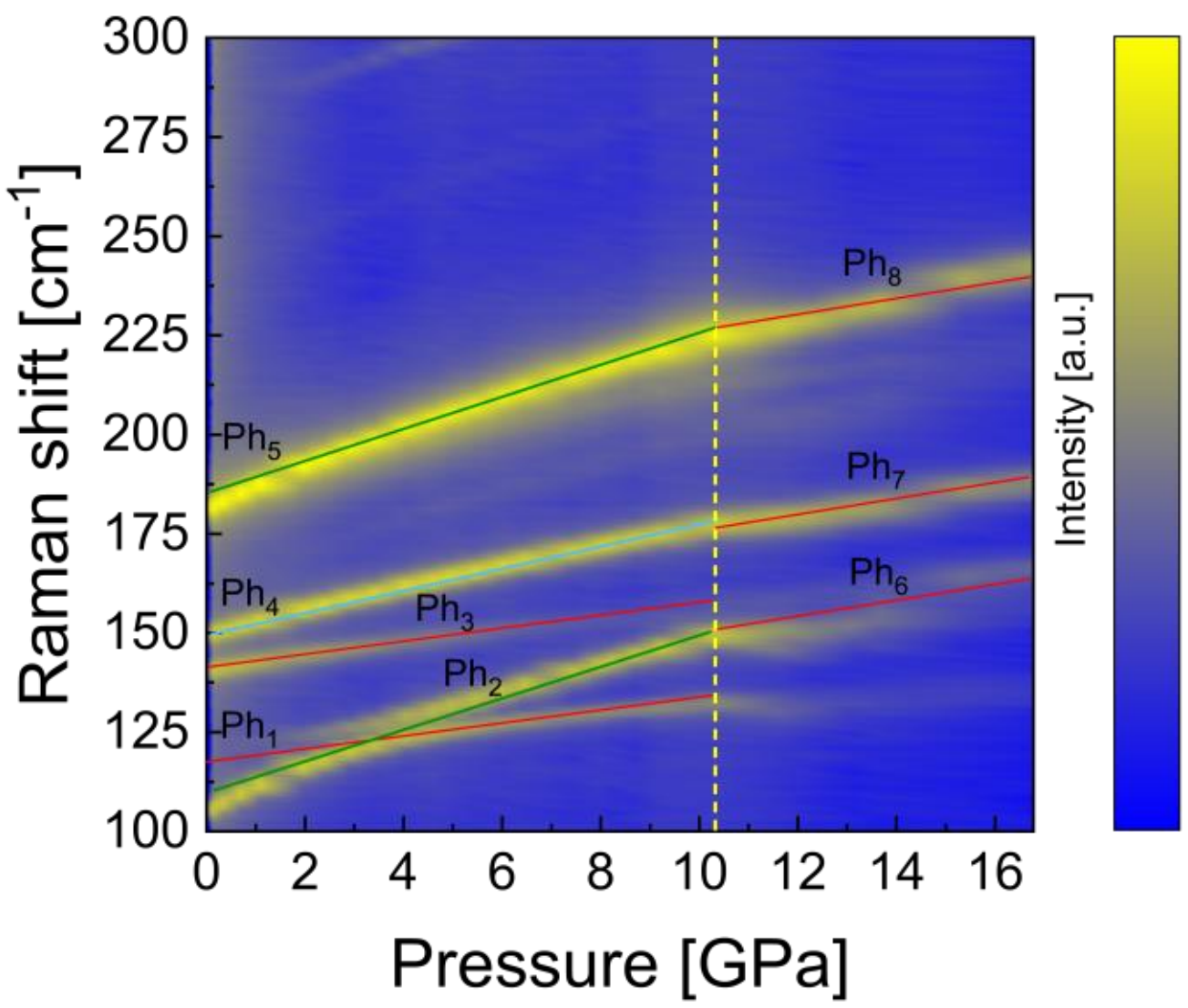

**Fig. 5 | The room temperature Raman spectrum map as a function of pressure.** The dashed line tentatively marks the $p_c$. Colored lines guide the eye to distinguish the different softness of the phonon modes under pressure. The red-marked phonon modes are assigned to $P\overline{3}m1$, the green ones (blue) to $R\overline{3}$ phase.

***Model of magnetic behavior under pressure***

Our density functional theory (DFT) calculations reproduce well the experimentally observed evolution of both lattice parameters under applied pressure (see Fig. 2e). The predicted density of states is shown in Fig. S6 [32]. An ambient-pressure gap of 1.67 eV was calculated, a value similar to that predicted for a monolayer[46]. Under the applied pressure of 10 GPa, the electronic structure changes considerably, and the gap shrinks to 1.39 eV (Fig. S6 [32]). This trend is expected if the material undergoes the suggested metallization at pressures above 20 GPa[12].

The calculated Cr-Br-Cr angle decreases from 94˚ at ambient conditions to the ideal value of 90˚ near 10 GPa (Fig. 2). Goodenough-Kanamori-Anderson rules predict that the superexchange with a 90˚ angle on the path via ligand is optimal for FM interaction, therefore, one may expect the nearest-neighbor exchange interaction to grow within this interval, as long as no transition to another structure occurs. On the other hand, direct Cr-Cr hopping prefers the AF orientation here and could become important as the lattice is compressed. DFT calculations could provide the answer to which of these competing effects is dominant.

Therefore, we evaluate exchange interactions by fitting energies of calculated supercells to the Heisenberg Hamiltonian. The fits to the neutron scattering data show that the second and third-nearest-neighbor interactions $J_2$, and $J_3$ are tiny, less than 10% of the nearest-neighbor $J_1$.[47] We disregard their pressure dependence and employ their experimental value. The calculated $J_1$ shows an initial slight increase with pressure, followed by a faster decrease (Fig. 6a). Our calculations indicate that intralayer FM coupling is preferred in this system at pressures up to 8 GPa. Critical temperatures are also affected by magnetic anisotropy, especially in monolayer or quasi-2D magnets. Therefore, we have evaluated the evolution of magnetic anisotropy energy with pressure (see methods). Within the interval of 8 GPa, it exhibits a factor of two decrease (Fig. S7), in approximate agreement with calculations showing it as a function of strain [15].

As the next step, we have performed finite-temperature Monte Carlo (MC) simulations using the calculated $J$'s and anisotropies at different pressures, assuming that the structure remains unchanged and only the ABC stacking is present. In this case, the $T_C$ exhibits a slight initial increase followed by a decrease; however, even at 8 GPa it does not fall below its ambient-pressure value (Fig. 6b). This behavior provides clear evidence that the description based on a single-phase is insufficient.

Recent work[25] examined the dependence of total energy and interlayer interactions in bilayer $CrBr_3$ on the layers' mutual position, finding that the magnetic exchange strength can be significantly varied, even changing sign. The AA stacking is precisely the case that favors the AFM coupling (Fig.1e, f in [25]) and is also most likely the realization of the layer arrangement observed in our high-pressure study. The above arguments suggest that phases with different stacking are present simultaneously in actual samples, and their relative proportions depend on the pressure. We denote the phase with standard stacking as the ABC phase, and AA stacking as the AA phase. Therefore, the system comprises a mixture of layers with (FM) in ABC phase. and AFM (in AA phase) interlayer interactions. The tendency to form different stable phases with different magnetic properties has also been observed in other vdW systems, namely $VI_3$[48].

To examine the effect of the simultaneous presence of different phases on the $T_C$, we have performed finite-temperature MC simulations for systems with varying concentrations of atoms in both AA and ABC phases. In vdW layered systems, different stackings affect properties of atoms only weakly due to the large separation between them, their effect on intralayer hopping terms, and thus the intralayer exchange can be neglected. The quantity most susceptible to the stacking order is the interlayer coupling, as the hopping paths and interatomic distances are changing.

Therefore, we assume that the phases differ only in the strength of interlayer interactions, and atoms from these phases are randomly distributed. Interlayer interaction $J_L$ corresponds here to a significantly higher energy per atom than MAE, hence reorientation unfavorable for the MAE requires much less energy than that which would align layers antiparallel (for ABC stacking), and one can expect a significant effect of $J_L$ on magnetic order.

The effective interlayer exchange was evaluated from our DFT calculations for $CrBr_3$ with either the ideal ABC or AA stacking, as shown in Fig. 6a. In our effective model we assume that the interlayer exchange depends only on the relative stacking of the two layers; the effect of stacking between more distant layers can be neglected. There may be also defects within the layers, where the layer geometry would be perturbed to accommodate a shift related to matching different phases (boundary of a stacking phase). Description of such situation in DFT calculation would require a very big supercell, and the results would depend a lot on the precise geometry of the boundary, which is not known. However, we can expect that there is only a very small number of atoms affected directly by these boundaries, compared to atoms more distant from them. Stacking changes in a vdW system occur primarily between layers. Therefore, we evaluate interlayer exchange couplings in ideal phases and assume their validity even when the surroundings of the studied layer pair are random. While the ABC stacked layers favor parallel interlayer alignment, those with AA stacking prefer the antiparallel one in the relevant pressure range above 2 GPa, similarly to what was found for $CrBr_3$ bilayers [25, 49]. For the concentration-dependent simulation, we have used exchange parameters predicted at 5 GPa, which corresponds to the state close to the loss of magnetic order. Our simulations reveal an intriguing trend: starting from the fully FM interlayer coupling (phase ABC only), the FM order decreases as the concentration of the AA phase grows, as shown in Fig. S8 [32]. The Curie temperature $T_C$ determined from the susceptibility exhibits a continuous decrease from the 39 K seen for the pure ABC phase – Fig. S8 [32]. This decline of $T_C$ with the AA phase concentration correlates with the change observed in pressure measurements.

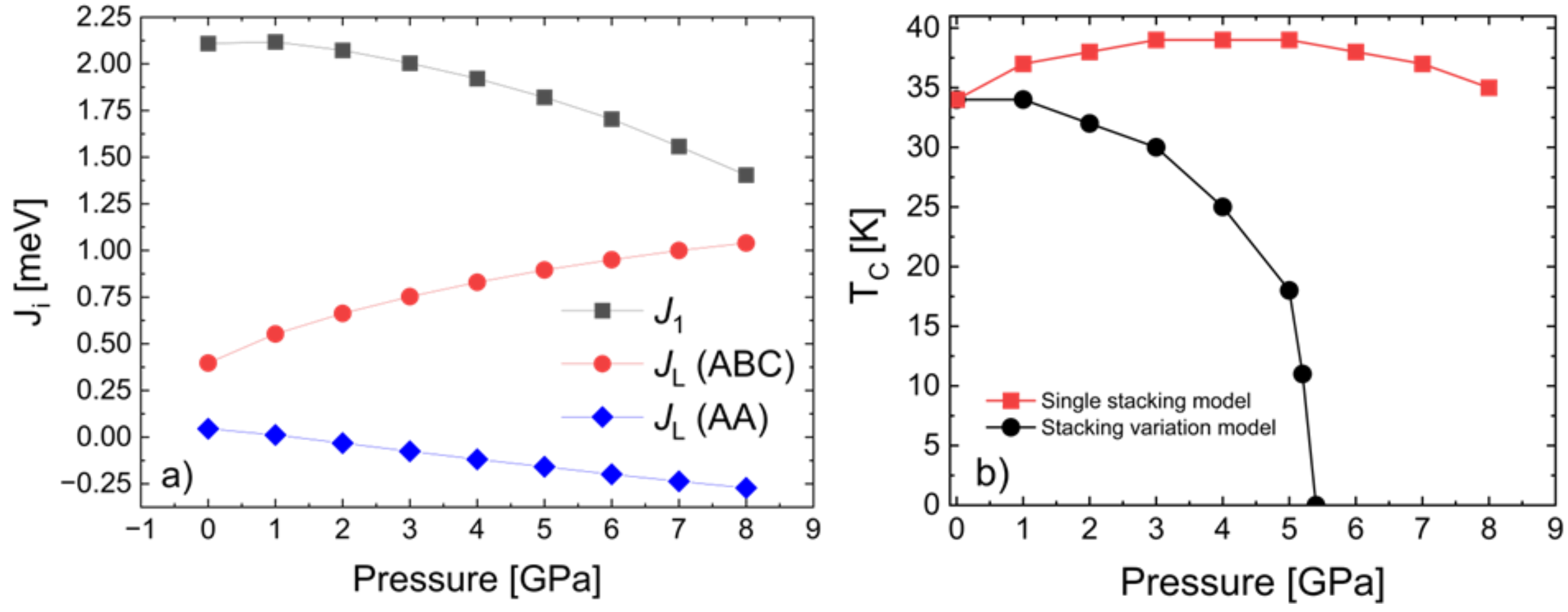

**Fig. 6 | Calculation results for bulk $CrBr_3$. a** The nearest neighbor interaction $J_1$ and interlayer interaction $J_L$ as a function of pressure (for both ABC and AA stacking). **b** The predicted Curie temperature $T_C$ as a function of pressure, obtained from atomistic simulations.

To connect the change in phase concentration to the dependence of exchange interactions, information about the pressure evolution of the structure is needed. This would allow us to estimate the $T_C(p)$ curve. From the current experimental data, we cannot describe the phase concentration with sufficient accuracy nor describe the exact structural arrangement of stacking variants. The assumption of random distribution is a standard modeling strategy in this situation. For the simulation of $T_C(p)$ curve, we describe the fraction of the AA phase by a linearly increasing function up to a critical pressure (further denoted as the stacking variation model), a similar approach that has been adopted for pressure-induced stacking faults in Gd [50]. The resulting dependence (Fig. 6b) allowed us to capture the key trend observed experimentally, the monotonic suppression of $T_C$ as the AA stacking fraction increases, which aligns well with the pressure-dependent magnetization data. The $T_C$ evolution obtained from the stacking variation model is much closer to the experimental data than the results from the ABC only stacking model (both in Fig. 6b).

The results of X-ray diffraction experiments and theoretical calculations provide a scenario that can consistently explain the seemingly controversial behavior of $CrBr_3$. The dominant ambient-pressure rhombohedral $R\overline{3}$ phase gradually decomposes with increasing pressure, accompanied by the formation of a layered structure favoring AFM coupling [25]. The ambient-pressure collinear FM ordering is gradually disturbed by the formation of AFM-coupled AA layers, which replace the original FM-coupled ABC layers. This process gradually decreases the Curie temperature and bulk magnetic moment despite the individual Cr-ion moments remaining intact.

Mixing of FM and AFM interactions represents a direct competition that leads to $T_C$ scaling with the weighted average of interactions, which differ not only in sign but also in magnitude. Finite-size effects will become important as fluctuations grow near criticality, causing deviations from this simple dependence. Unlike mean field approximation, Monte Carlo / spin dynamics simulations capture these, but only with limited accuracy [51]. The magnetic correlation length is truncated here by stacking faults, leading to a highly anisotropic way. It is constrained mainly in the z direction, where the interaction is also much weaker. $T_C$ is thus much less affected than it would be in the case of constriction in-plane.

There are other effects whose description in a simulation would be challenging. Stacking faults may promote domain formation. These were observed in $CrBr_3$ constrained to a monolayer in connection with increased fluctuations near Tc, where the long-distance dipole-dipole interactions also play a bigger role [26]. Magnetic order persists up to monolayer thickness with a $T_C$ of approx. 70% of the unconstrained bulk. Pressure affects both the volume fraction of the two stackings and the magnetic interactions. For all the above reasons, the relation between the critical temperature and the volume fraction of the two stackings is rather complex, and one cannot draw any simple conclusions about the suppression of one stacking at critical pressure.

A question remains about the character of the magnetic phase at pressures beyond $p_c$, when the localized ionic magnetic moments are preserved but the bulk spontaneous moment vanishes. The magnetic nature of this pressure-induced phase was the subject of the performed DFT calculations. The positive interlayer exchange constants for the ABC layer stacking were found, while the negative ones for the AA layer were found. The simulation results corroborate the trend of the fully FM interlayer coupling of the pure $R\overline{3}$ phase with ABC layer stacking to a state with 45% of $P\overline{3}m1$ phase with AA layer stacking, which favors the AFM coupling. At pressures higher than $p_c$, the $P\overline{3}m1$ phase with AFM-coupled layers dominates. The resulting magnetic structure is generally AFM with zero bulk moment. A microscopic experiment resolving details of the magnetic structure at pressures above $p_c$ is desirable to verify this scenario.

CONCLUSIONS

We have investigated the structural and magnetic response of $CrBr_3$ to external pressure. Ambient-pressure single-crystal diffraction revealed intrinsic trimorphism: the known high-temperature monoclinic and low-temperature rhombohedral $R\overline{3}$ phases, together with a so far unreported trigonal phase $P\overline{3}m1$. This phase is characterized by AA stacking and 2/3 occupation of the Cr $1a$ site. Despite coexistence with $R\overline{3}$ at ambient conditions, high-resolution reciprocal lattice maps have shown the layers are ideally spaced along the *c*-axis, irrespective of structure, and the mixed stacking of the structural blocks is well described by a paracrystal model.

High-pressure diffraction showed that $P\overline{3}m1$ becomes the single stable phase at the highest pressures and, notably, persists on full decompression from 21.5 GPa. This result suggests that the $P\overline{3}m1$ phase has been stabilized at the highest pressure and preserved during decompression to ambient pressure, where it remains metastable.

We provide the first direct experimental evidence of pressure-induced suppression of ferromagnetism in $CrBr_3$. Both the Curie temperature and saturated magnetic moment at 2 K decrease continuously with pressure, with accelerated suppression above ~3 GPa and a critical pressure of $p_c \approx 5.8$, significantly lower than prior extrapolations reported in literature. The loss of ferromagnetism is consistently explained by pressure-driven changes in layer stacking that introduce interlayer AFM interactions, as predicted by a combination of DFT calculations and atomistic finite-temperature magnetism simulations. The ferromagnetic phase fully recovers after decompression from 10 GPa to ambient pressure, indicating that the $P\overline{3}m1$ phase was not yet stabilized at 10 GPa while some volume fraction of $R\overline{3}$ persisted. Microscopic magnetic probes at high pressures above $p_c$ will be essential to resolve the nature of the emerging high-pressure magnetic phase.

## EXPERIMENTAL METHODS

### *Sample synthesis*

The single crystals of $CrBr_3$ have been grown directly from pure elements (Cr 99.99 %, $Br_2$ 99.5%) using the chemical vapor transport method. We have invented the identical procedure developed previously for single crystal growth of $VBr_3$[52]. A thermal gradient of 700/600°C was kept for two weeks to transport all Cr metal from the hot part of the tube. The single crystals of the black-reflective color exceeding a cm-scale have been obtained. The desired 1:3 composition was confirmed by EDX analysis. The crystallinity and orientation of the single crystals were confirmed by the Laue method, which showed sharp reflections. The rhombohedral *c*-axis is perpendicular to the plane of plate-like crystals.

### *X-ray diffraction methods at ambient and high pressures*

The samples were measured on a Rigaku OD Gemini four-circle diffractometer equipped with Atlas S2 CCD detector, using graphite monochromated Mo-$K\alpha$ radiation ($\lambda$ = 0.71073 Å) from a sealed X-ray tube. The samples were measured at room temperature. The data collection was handled by CrysAlis PRO[53]. In samples measured at ambient pressure, the measured diffractions were integrated, scaled, and corrected for absorption using CrysAlis PRO [53]. The initial structure models of $CrBr_3$ were solved using charge flipping in Superflip[54] and refined by full-matrix least-squares on $F^2$ in Jana2020 [55]. For high-pressure measurements, the sample was enclosed in the Merryll-Bassett type diamond anvil cell (ALMAX–Easylab bv, https://www.almax-easylab.com/) with an 85° opening angle. The data collection strategy was planned in CrysAlis PRO [53], considering the limitations imposed by the diamond anvil cell. The Ruby fluorescence line was used for pressure calibration with a liquid pressure medium.

Comparative X-ray powder diffraction experiments at ambient pressure were performed on a Bruker D8 Advance X-ray diffractometer using Cu-$K\alpha$ radiation and working in the Bragg-Brentano parafocusing geometry. The acquired data were evaluated by using the FullProf program. The single crystals installed inside the cells are displayed in Figs. S9-S11 [32].

RIGAKU Smartlab diffractometer using 45kV/200mA power Cu-$K\alpha$ radiation, primary channel-cut monochromator, and a one-dimensional detector were used to measure high-resolution reciprocal-space maps around chosen reciprocal lattice points. In order to investigate also the structural changes due to pressure, we measured X-ray diffraction reciprocal space maps around the reciprocal lattice points 10-1 13 and 30-3 15 before and after deformation by external force. For the measurements, we used a high-resolution setup with a parabolic multilayer mirror and a 2x220Ge channel-cut monochromator on the primary side, and a 0.02rad Soller slit and a two-dimensional detector on the secondary side.

In the course of preparation for the high-pressure single crystal diffraction experiments on $CrBr_3$, we encountered an unexpected phenomenon. During the initial experiments, we found that samples that showed the correct $R\overline{3}$ structure, when measured “as-grown”, demonstrated the presence of the $P\overline{3}m1$ phase already at ambient pressure after being cut down to a suitable size for the DAC experiment. We have tested a large number of samples (30+) and found the presence of a portion of the $P\overline{3}m1$ structure in each case. It leads us to conclude that the $CrBr_3$ layered structure is highly delicate and prone to physical incursions, such as cutting, bending, exfoliating, etc. Selecting a suitable $R\overline{3}$ single-phase specimen for the high-pressure study was somewhat of a matter of chance. Finally, the best sample used in the end showed ~80% volume fraction of the correct ambient pressure $R\overline{3}$ rhombohedral phase at the start of the high-pressure experiment.

***High-pressure magnetization***

The magnetic properties of $CrBr_3$ under high pressures have been probed by direct magnetization measurements in a SQUID magnetometer (Quantum Design, Ltd.), using the miniature turnbuckle-type diamond anvil cell (DAC) made of ultrapure custom-made Be/Cu alloy [56] with a liquid pressure medium and a ruby sphere (Almax) as a manometer. Reference measurement of the empty cell without a sample has been used to subtract the background signal of the pressure cell, ruby, and pressure-transmitting medium. Some reference ambient pressure measurements have been done using the VSM option of the PPMS14 apparatus (Quantum Design).

***Raman spectroscopy under high pressures***

The polarized Raman spectra were recorded in a back-scattering geometry using an RM1000 Renishaw micro-Raman spectrometer with the 514.5 nm line of an $Ar^+$ ion laser at a power of about 1 mW (~100 μW on the sample) in the spectral range 10–540 $cm^{-1}$. The spectrometer was equipped with Bragg grating filters, which enabled good stray light rejection. The diameter of the laser spot on the sample surface was ∼5μm, and the spectral resolution was better than 1.5 $cm^{-1}$. For high-pressure measurements at room temperature, the sample was enclosed in the Merryll-Bassett type diamond anvil cell equipped with type II as diamonds (ALMAX –Easylab bv, https://www.almax-easylab.com/), liquid pressure transmitting medium, and a ruby pressure scale.

***General comment on pressure experiments***

In all types of DAC experiments and different measurement techniques, the pressure was loaded at room temperature and determined using the ruby pressure scale [57, 58]. Ruby spheres from Betsa [59] has been used in all pressure experiments.

***Ab initio calculations***

We have utilized density functional theory (DFT) calculations to describe the electronic structure. Ionic positions of bulk $CrBr_3$ were optimized using the Vienna Ab Initio Simulation Package (VASP)[60]within the framework of the projector augmented wave (PAW) method [61]. The exchange-correlation functional was treated using the generalized gradient approximation (GGA) in the Perdew-Burke-Ernzerhof (PBE) scheme [62]. To account for the interlayer long-range van der Waals (vdW) interactions, the non-local vdW functional was incorporated via the DFT-D3 approach[63, 64]. A plane-wave energy cutoff of 520 eV was employed, and the Brillouin zone was sampled using a 7×7×3 Monkhorst-Pack k-point grid [65].

The optimization involved relaxing all lattice parameters and ionic coordinates until the forces acting on each ion were reduced to below $10^{-2}$ eV/Å. On-site Coulomb interactions were incorporated through the GGA+U method, with an effective U parameter ($U_{eff}$) of 2.8 eV to better describe the localized d-electrons of Cr atoms [66, 67]. Stress tensor optimization allowed for the simultaneous relaxation of the lattice and atomic positions under external pressure, with pressure effects simulated from 1 GPa to 10 GPa. Density of states (DOS) calculations were performed using Gaussian smearing to achieve high accuracy. For the calculation of exchange interactions we have adopted the $r^2$SCAN meta GGA approximation together with the nonlocal rVV10 van der Waals (vdW) functional [64, 68, 69]. This method has been shown to obtain more accurate total energies than many other combinations of exchange-correlation and vdW functionals for hexagonal layered solids (on average), as studied on a set of 10 materials [70].

Magnetocrystalline anisotropy energy (MAE) calculations were calculated with spin-orbit coupling (SOC) within the noncollinear compilation of VASP as the difference between total energies corresponding to the magnetization in the in-plane and out-of-plane directions (MAE = $E_{\parallel} - E_{\perp}$). Therefore, a positive MAE value indicates easy axis. SOC calculations were converged with the self-consistent field (SCF) approach with strict convergence criteria (difference in consecutive SCF steps less than $10^{-7}$ eV, or $10^{-6}$ eV where the former was not feasible. To get precise total energies, the final SOC calculations were performed using the tetrahedron approach with Blöchl adjustments [71].

The effective interlayer exchange was evaluated from our DFT calculations for $CrBr_3$ with either the ideal ABC or AA stacking. In our effective model we assume that the interlayer exchange depends only on the relative stacking of the two layers; the effect of relative stacking between more distant layers can be neglected.

***Atomistic simulations***

Finite-temperature magnetism is studied using the Uppsala Atomistic Spin Dynamics (UppASD) framework [51, 72, 73], which incorporates magnetic exchange energies from density functional theory (DFT) calculations. UppASD enables the simulation of spin dynamics and magnetic properties at interatomic scales, providing an accurate means to determine key parameters such as the Curie temperature ($T_C$). Thermodynamic properties are computed using a combination of Monte Carlo (MC) and spin dynamics methods.

To achieve equilibrium, a system of ten layers of 20×20×1unit cells evolved over 90,000 MC steps per temperature step under periodic in-plane and interlayer boundary conditions. The mean magnetization is then extracted using spin dynamics with an additional 90,000 steps per temperature. Three parallel simulations are performed for each configuration to improve statistical accuracy, and the results are averaged.

The spin dynamics simulations solve the Landau-Lifshitz-Gilbert (LLG) equation, which describes spin evolution by accounting for precession and damping effects. Numerical integration is performed with a time step of $\Delta t = 0.1$ femtoseconds ($10^{-16}$ s). At finite temperatures, random microscopic scattering events are modeled via a stochastically fluctuating thermal field, consistent with the Langevin dynamic [51] and the fluctuation-dissipation theorem, linking the field variance to temperature and damping. The MC Metropolis algorithm is used and adapted for spin dynamics to enhance computational efficiency for time-independent magnetization dynamics.

A variation of the Heisenberg Hamiltonian $H$ concerning small changes in $m_i$ is evaluated to calculate the effective field around each magnetic ion. In layered $CrBr_3$, the Hamiltonian incorporates intralayer ($H \parallel$), interlayer ($H \perp$), and single-ion anisotropy ($H_{SIA}$) terms:

Effective Hamiltonian:

$$H = H_{\parallel} + H_{\perp} + H_{SIA} = \frac{-1}{2}\sum_l \sum_{i,j} j_{ij} \hat{e}_{l,i} . \hat{e}_{l,j} - \sum_l \sum_{i,j} j'_{ij} \hat{e}_{l,i} . \hat{e}_{l+1,j} + \sum_l \sum_i E_{SIA}(\hat{e}_{l,i}) \quad (1)$$

where $\hat{e}_{l,i}$ represents the spin direction of the ith magnetic ion in the l[th] layer, $j_{ij}$, and $j_{ij}'$ are intralayer and interlayer exchange parameters, and $E_{SIA}$ is the single-ion magnetic anisotropy energy. One can also define an effective interlayer interaction $J_L = \frac{1}{N_{2D}}\sum_{i,j} j'_{ij}$, where $N_{2D}$ is the total number of magnetic ions in the layer. Its value can be directly extracted from DFT calculations by comparing the energy of the FM state $E_{FM}$ and the layered AFM state $E_{AFL}$, noticing that $E_{AFL} - E_{FM} = 2.N.J_L$ , where N is the total number of magnetic ions in the calculation cell. This comprehensive approach captures critical magnetic interactions and stochastic effects, enabling accurate modeling of finite-temperature magnetism in layered ferromagnetic systems.

**Acknowledgments**

This work is a part of the research project 25-15448S funded by the Czech Science Foundation. Experiments were performed in MGML (mgml.eu), supported within the Czech Research Infrastructures program (project no. LM2023065). The authors also acknowledge the assistance provided by Project TERAFIT - CZ.02.01.01/00/22_008/0004594, co-financed by the European Union and the Czech Ministry of Education, Youth and Sports. KP is supported by the Grant Agency of Charles University (GA UK), Project No. 155424.

**Additional information**

The online version contains supporting material.

**Data availability**

Data are available upon request.

**Code availability**

This study did not involve the development of any custom code or novel mathematical algorithms.

**Author Contribution**

M. M., U. D., P. K., J. K., J. V. J. Pr., J. K., and J. V. performed the pressure experiments of various physical quantities, V. E., M. D., V. H. measured and analyzed X-ray diffraction data, F. B. and S.K. measured and analyzed the spectroscopy data, K. P., S. R., and K. C. provided the theoretical calculations, J. Po. and D. H. has grown the single crystals, V. S. and J. P. coordinated work and wrote the manuscript, all coauthors revised the manuscript before the submission.

**Competing interests**

The authors declare no competing interests.

Correspondence and requests for materials should be addressed to Jiří Pospíšil (jiri.pospisil@matfyz.cuni.cz) or Martin Míšek (misek@fzu.cz)

# Supporting Information for:

## Pressure-induced structure transformation and collapse of ferromagnetism in van der Waals insulator

*M. Míšek[1], U. Dutta[1], P. Král[2], D. Hovančík[2], J. Kaštil[1], K. Pokhrel[2], S. Ray[2], J. Valenta[1,2] J. Prchal[2], J. Kamarád[1], F. Borodavka[3], V. Eigner[1], M. Dušek[1], V. Holý[2], K. Carva[2], S. Kamba[3], V. Sechovský[2] and J. Pospíšil[2]*

[1]Institute of Physics of the Czech Academy of Sciences, Cukrovarnická 10/112, 162 00 Prague 6, Czech Republic

[2] Charles University, Department of Condensed Matter Physics, Ke Karlovu 5, 121 16 Prague 2, Czech Republic

[3]Institute of Physics of the Czech Academy of Sciences, Na Slovance 2, 182 00 Prague 8, Czech Republic

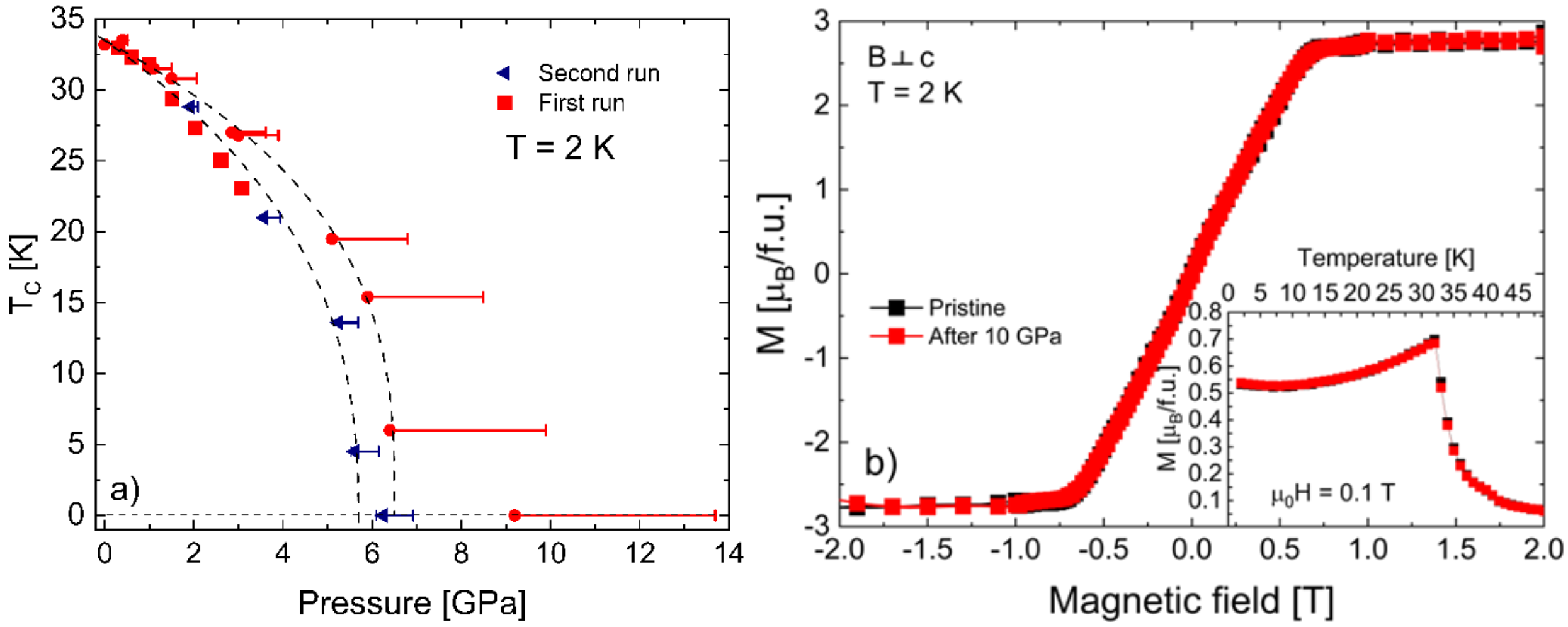

**Fig. S1 | a** The magnetic *p*-*T* phase diagram of $CrBr_3$, in which both sets at of measurements are shown. The value of $p_c$ where the ferromagnetism is suppressed deviates by about 1 GPa. The horizontal right-side error bars demonstrate the pressure value inside the cell before the moment when a new stable pressure point is set. **b** Magnetization loops measured at ambient pressure before pressure loading and after the decompression from 10 GPa; the inset shows the corresponding temperature dependences of magnetization measured in 0.1 T.

The construction of the turnbuckle-type DAC for the SQUID measurements requires that the target pressure has to be exceeded during loading, as the cell relaxes after being removed from the press/loading frame. The original design of the DAC aimed for minimizing the body of the cell and consequently the background signal. During the course of $CrBr_3$ experiments, we realized this overshoot can be quite significant and reworked the design of the pressure cell completely. For the 2nd experiment, we have built a new cell with a more massive body to minimize the pressure drop during the loading process.

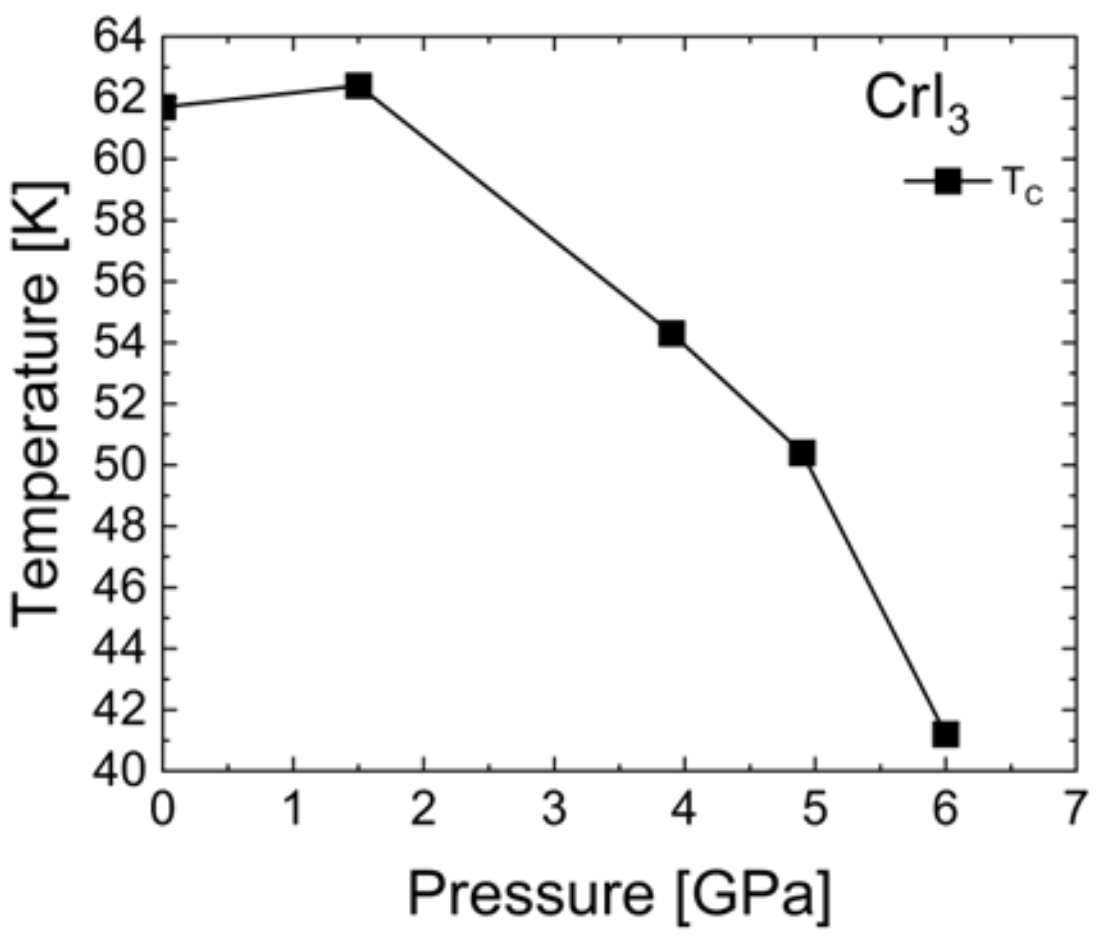

**Fig. S2 |** Pressure evolution of the Curie temperature in $CrI_3$ as a function of increasing hydrostatic pressure. Data are in agreement with previous work[1].

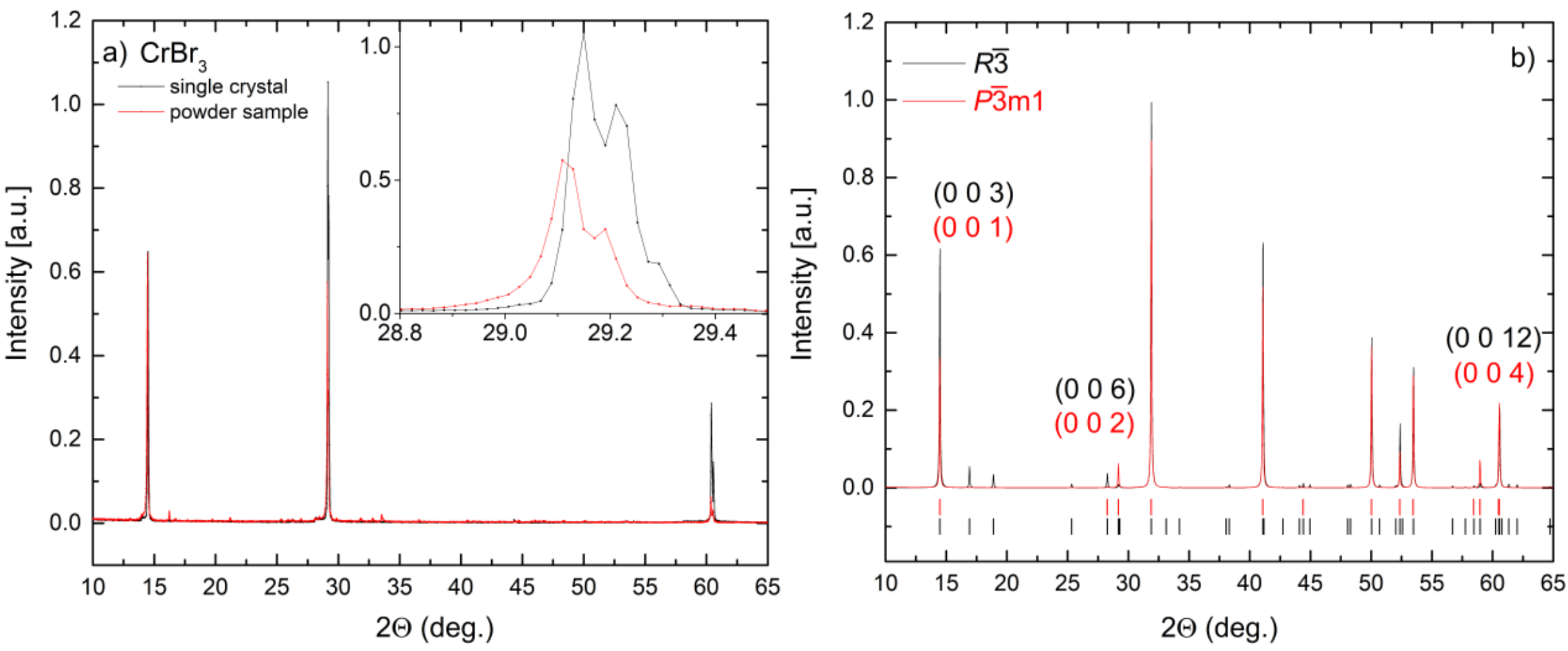

**Fig. S3 | a** The real and **b** simulated X-ray powder diffraction patterns of both $CrBr_3$ structural phases.

Since the substantial part of the work reported on the bulk $CrBr_3$ samples of the trihalide family was done using powder X-ray and neutron diffraction methods [2-4], it was of natural concern how the milling into the powder form affects the $CrBr_3$ sample in terms of both structural and magnetic properties. We ground the CVT-grown single crystals in powder form and subjected them to the X-ray powder diffraction experiment. The result is shown in Fig. S3, which includes the theoretical X-ray powder diffraction patterns of both structures modelled based on single crystal diffraction data in Tables S1-4. It is evident that $R\overline{3}$ and $P\overline{3}m1$ are due to the almost ideal overlap of the most intensive reflection indistinguishable by this method; thus $P\overline{3}m1$ phase remained unresolved in studies in the past.

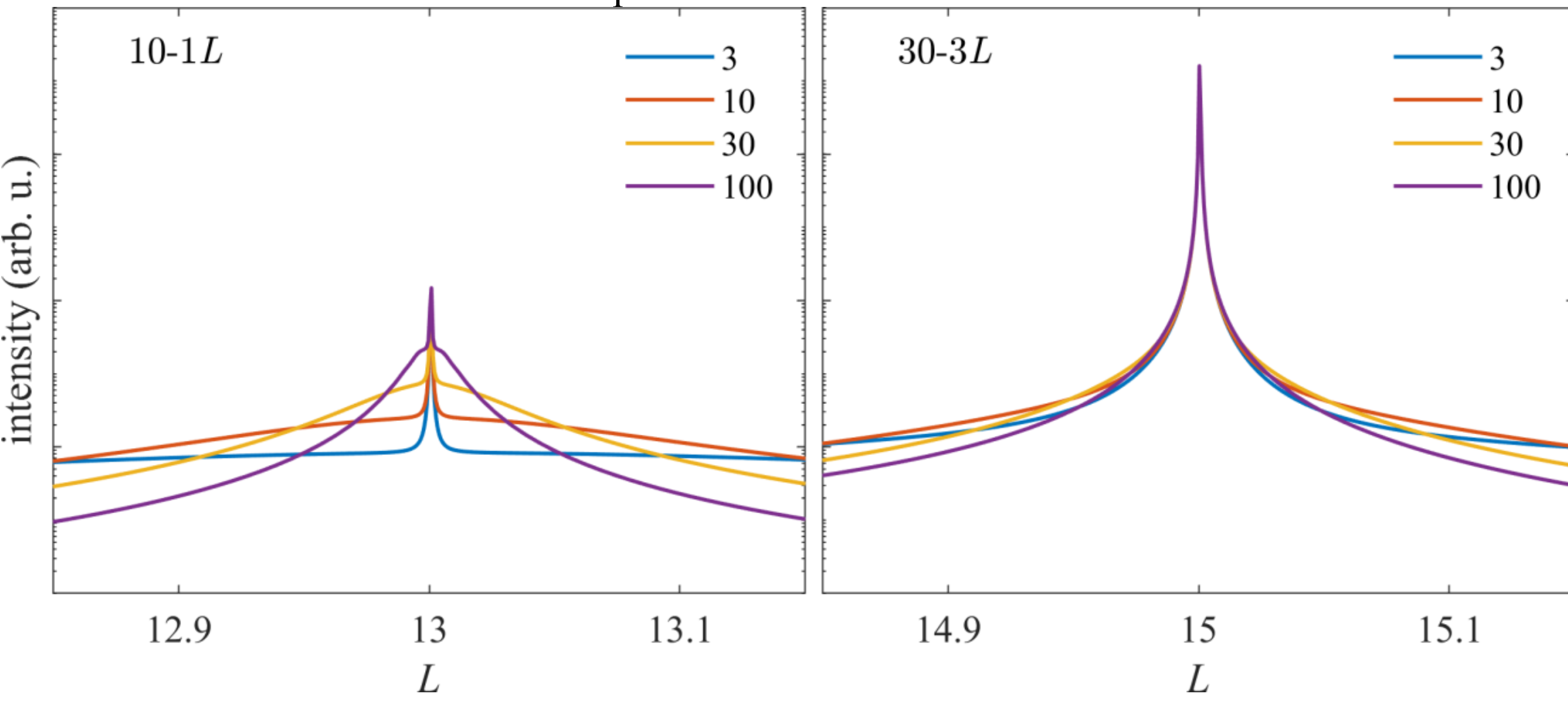

**Fig. S4 | Simulations of the paracrystal model byHendricks-Teller model.** Diffraction curves of the 10-1 13 (left) and 30-3 15 diffraction maxima (right) calculated for various mean block lengths; the block lengths are distributed according to a Poisson distribution. The parameter of the curves is the mean block length in number of monolayers.

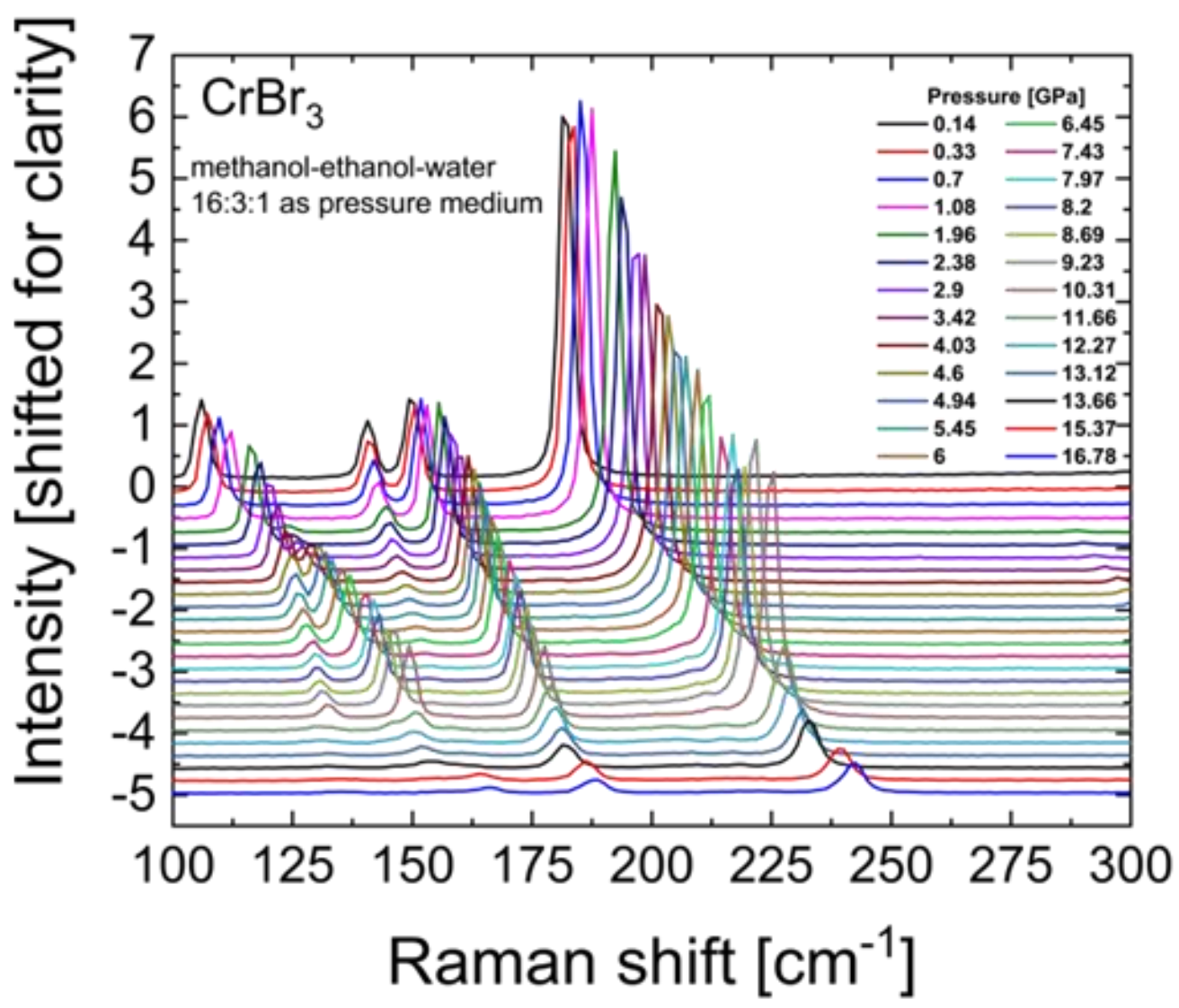

**Fig. S5 |** The room temperature Raman spectra at all measured pressures with a constant offset in the intensity scale for better data readability.

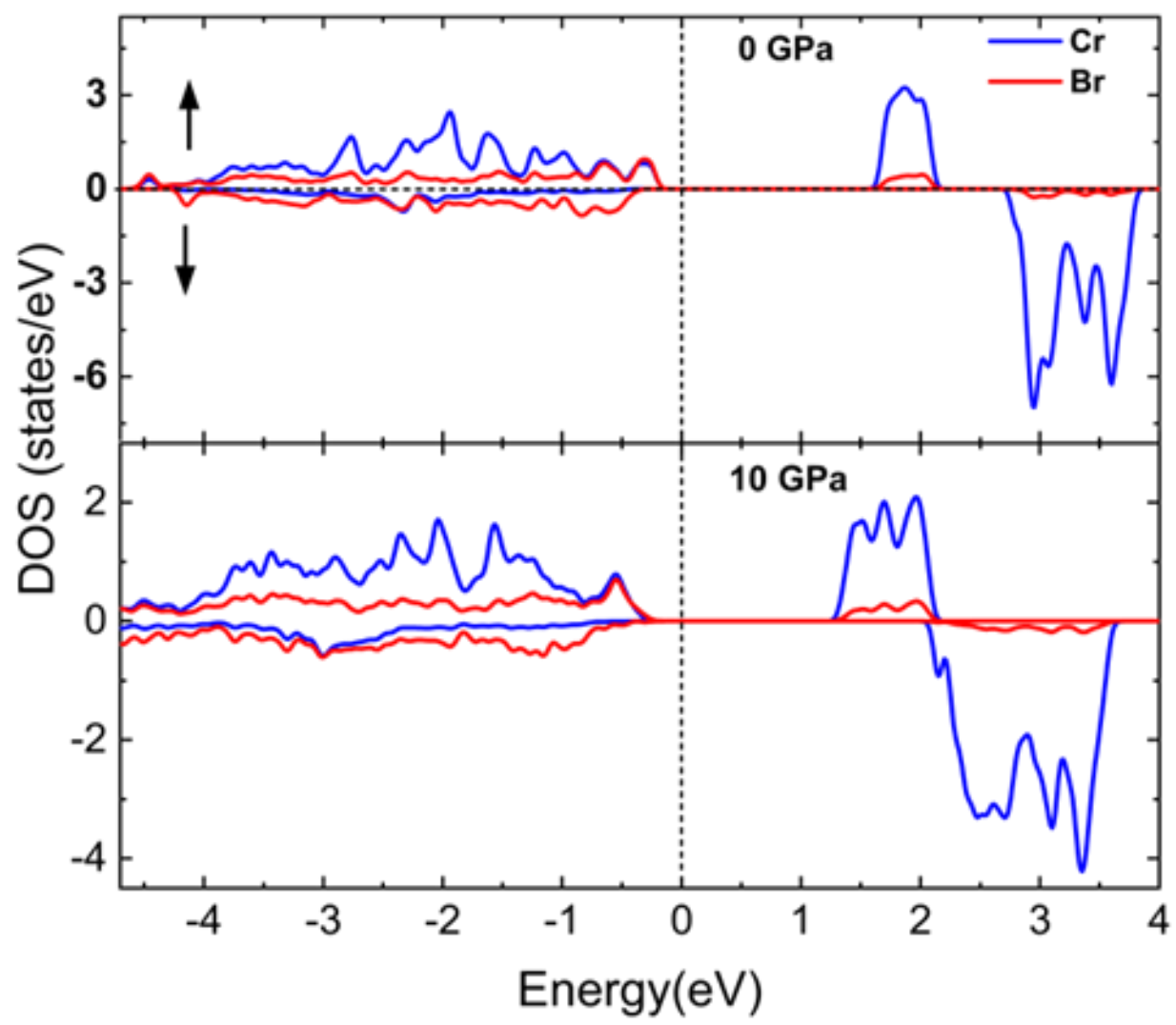

**Fig. S6 |** The density of states of $CrBr_3$ for ambient pressure and 10 GPa.

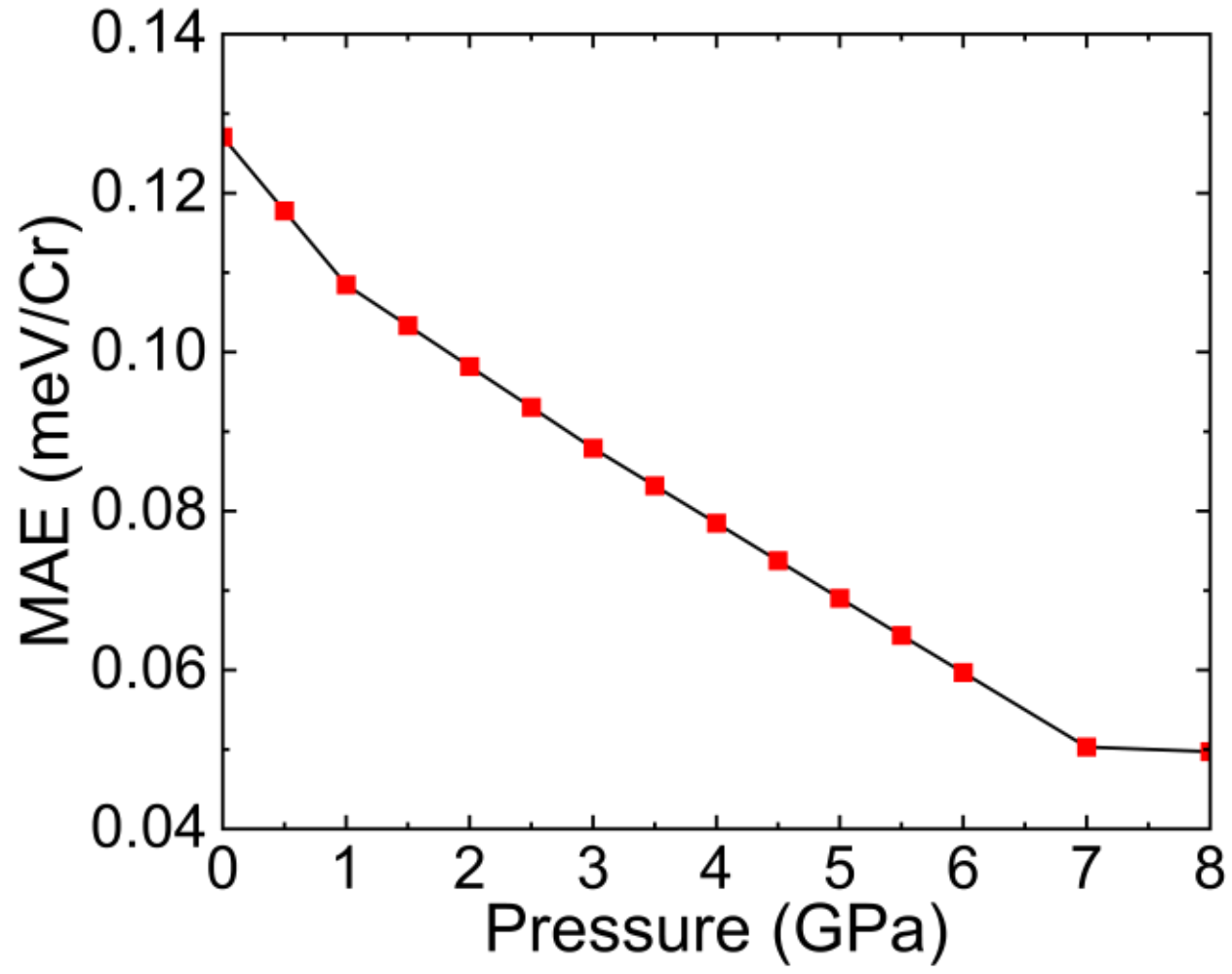

**Fig. S7 |** Calculated magnetic anisotropy energy (MAE) of $CrBr_3$ as a function of pressure.

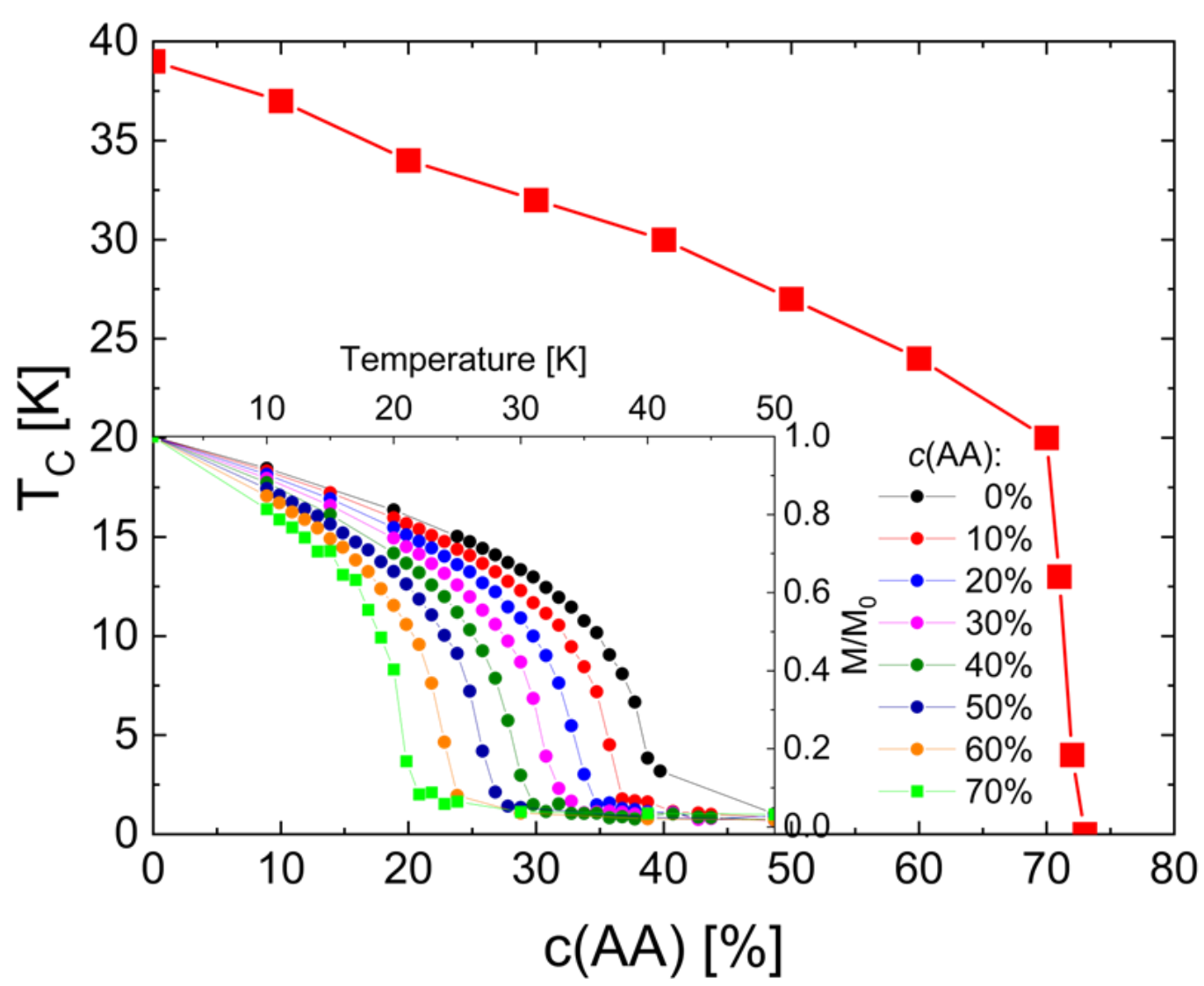

**Fig. S8 |** Monte Carlo simulation results: Dependences of the Curie temperatures $T_C$ for bulk $CrBr_3$ on the AA phase concentration, assuming 5 GPa pressure values of exchange interactions. Inset: Simulated M(T) curves for various AA phase concentrations c(AA).

Temperature-dependent magnetization was calculated using Monte Carlo simulations within the UppASD package. All simulations are performed using atomistic simulation method (see manuscript for more details) at fixed exchange interaction parameters obtained at 5 GPa from DFT. Fig. S7 shows that gradual increment in the concentration of AA-phase continuously decreases $T_C$ , which vanishes completely at 73% -AA concentration.

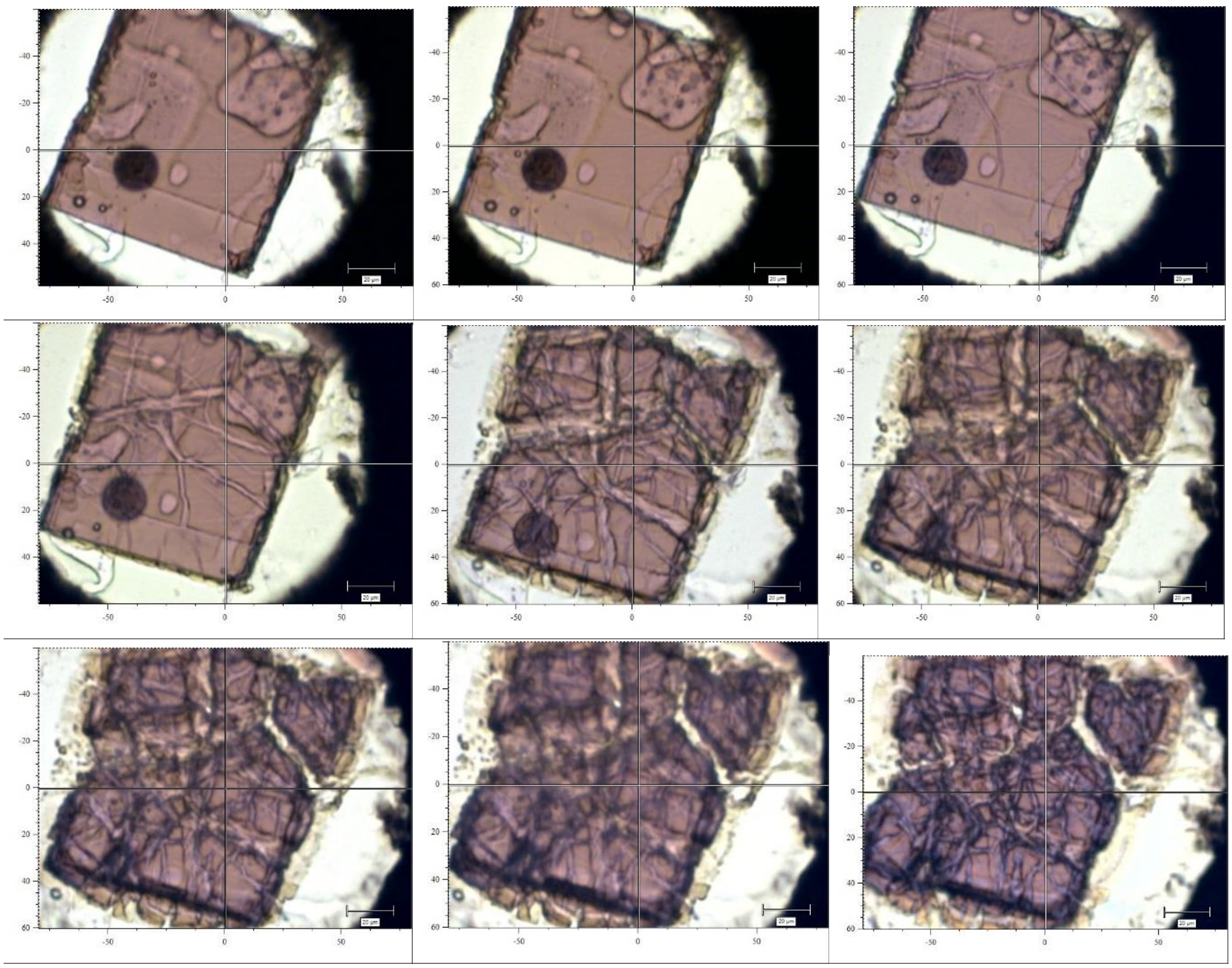

**Fig. S9 |** $CrBr_3$ sample pressurized using Daphne 7474 pressure medium at various pressures (top left ~1.2 GPa to bottom right ~17.1 GPa).

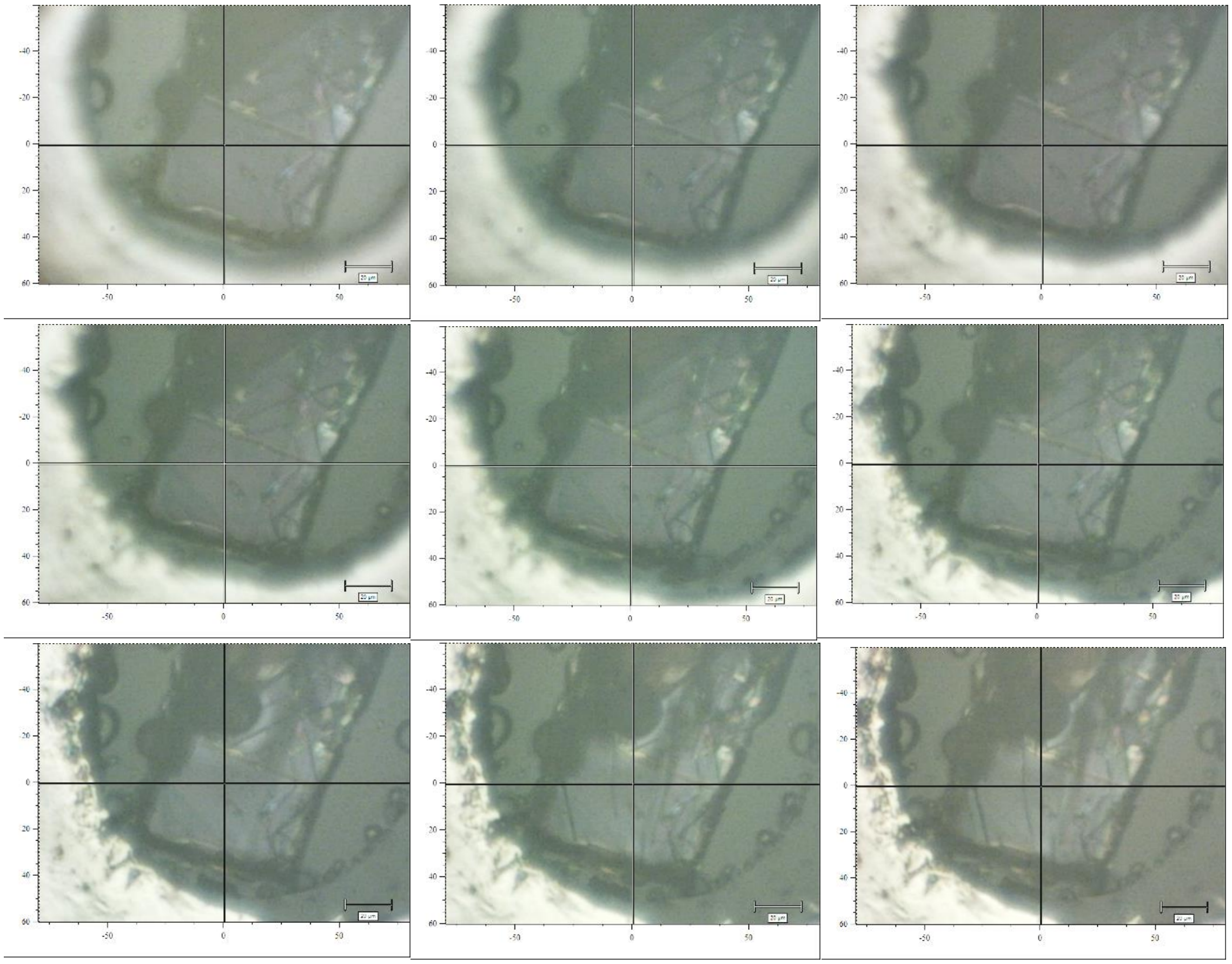

**Fig. S10 |** $CrBr_3$ sample compressed within the methanol-ethanol-water 16:3:1 mixture at various pressures. Top left – 0.7 GPa to bottom right – 13.12 GPa. These pictures were taken during the Raman experiment depicted in the contour plot in Figure 5 in the main paper. The photos were taken during the Raman experiment using the MEW medium, and the measured data are depicted in Fig. S5. Compared to the disintegration of the sample in non-hydrostatic media (Fig. S8), the sample stays almost intact to pressures well above 10 GPa.

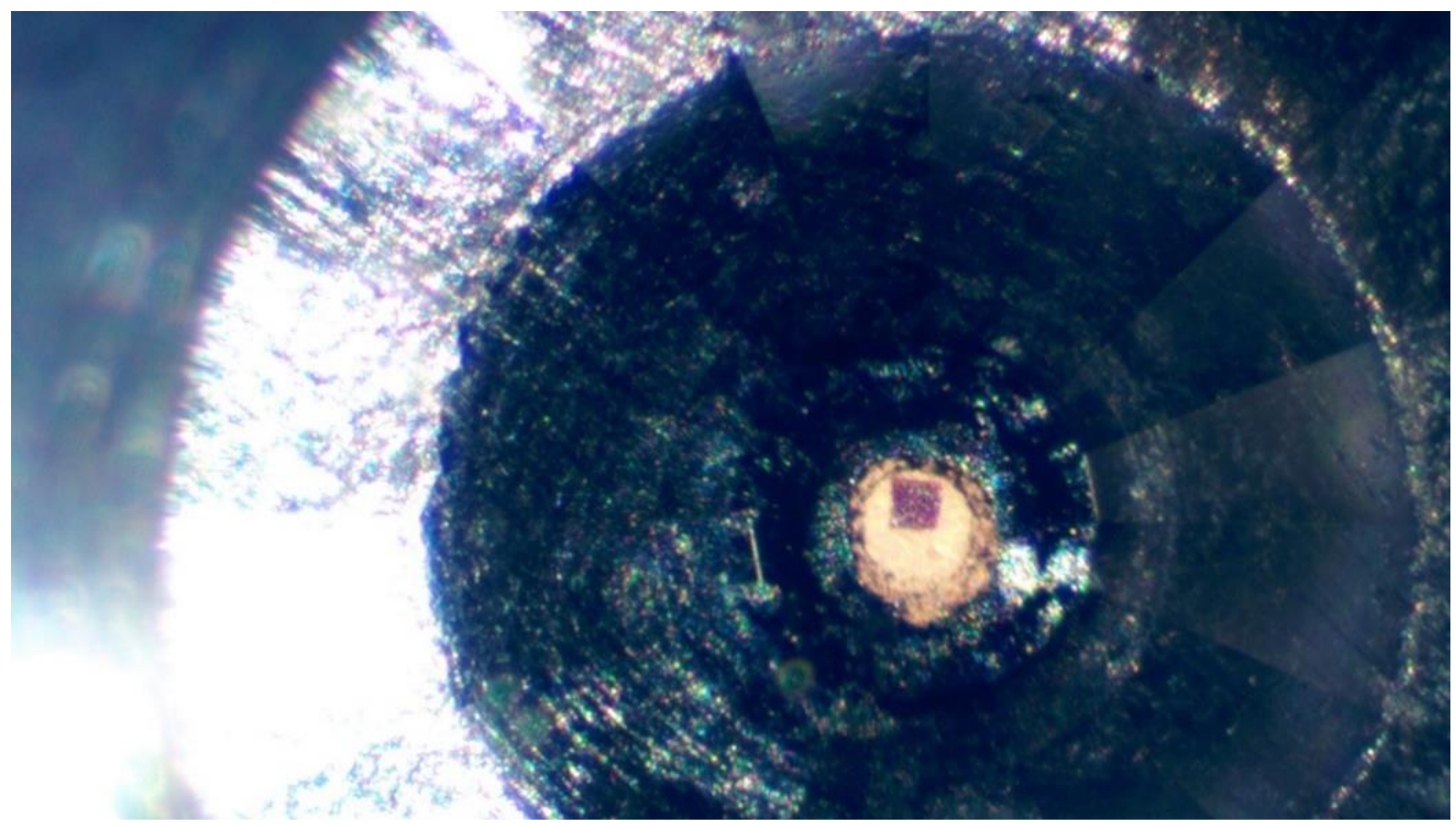

**Fig. S11 |** $CrBr_3$ sample pressurized in MEW medium to ~ 17 GPa, the picture was taken during the single crystal X-ray diffraction experiment.

***General remarks on the high-pressure environment***

The quality of the high-pressure conditions is essential, especially in materials like $CrBr_3$, which are extremely sensitive to minute physical stimuli. The presence of shear stress or inhomogeneous pressure distribution could seriously hamper the experimental effort. We tried various pressure media to load the DAC. Ultimately, all the experiments reported in this paper were prepared using the methanol-ethanol-water mixture (16:3:1) with a hydrostatic limit of ~10.5 GPa [6-8]. Given the known sensitivity of the other $MX_3$ compounds to water, samples were first tested by submersion in PTM for ~2 weeks, with no observable signs of degradation. To illustrate the seriousness of non-hydrostatic conditions in the studied system, a few photographs of the sample pressurized under different pressure-transmitting media have been added to Figs S8-S10.

**Table S1 | Crystallographic data for both crystal structures of $CrBr_3$ at 289 K.**

| | $R\overline{3}$ | $P\overline{3}m1$ |
|---|---|---|
| Formula | $CrBr_3$ | $Cr_{0.67}Br_2$ |
| $M_r$ | 291.7 | 194.5 |
| Crystal system | Rhombohedral | Trigonal |
| $a$ (Å) | 6.3091(7) | 3.6439(17) |
| $b$ (Å) | 6.3091(7) | 3.6439(17) |
| $c$ (Å) | 18.342(2) | 6.116(3) |
| $\alpha$ (°) | 90 | 90 |
| $\beta$ (°) | 90 | 90 |
| $\gamma$ (°) | 120 | 120 |
| $V$ (Å$^3$) | 632.27(12) | 70.33(6) |
| $Z$ | 6 | 1 |
| $F$(000) | 774 | 86 |
| Dimensions (mm) | 0.12 x 0.09 x 0.01 | 0.15 x 0.11x 0.02 |
| $\mu$ (mm$^{-1}$) | 31.00 | 30.77 |
| $\rho_{calc}$ (g cm$^{-3}$) | 4.597 | 4.592 |
| $T$ (K) | 289 | 296 |
| Radiation, $\lambda$ (Å) | Mo-$K\alpha$, 0.71073 | Mo-$K\alpha$, 0.71073 |
| $\theta_{min}$, $\theta_{max}$ (°) | 3.33, 29.11 | 3.33, 29.10 |
| Limiting indices | h= -8→8 | h= -4→4 |
| | k= -8→8 | k= -4→4 |
| | l= -24→24 | l= -8→7 |
| Collected reflections | 3203 | 356 |
| Independent reflections | 373 | 90 |
| $R_{int}$ | 0.0719 | 0.047 |
| Observed reflections | 260 | 78 |
| Refined parameters | 16 | 6 |
| Restraints | 0 | 0 |
| Constraints | 1 | 0 |
| $R_1$ [$I$>3$\sigma$($I$)] | 0.0429 | 0.0417 |
| $wR_2$ [$I$>3$\sigma$($I$)] | 0.1031 | 0.0977 |
| $R_1$ (all) | 0.0647 | 0.0494 |
| $wR_2$ (all) | 0.1153 | 0.0996 |
| Goodness of fit | 1.34 | 1.60 |
| $\Delta\rho_{min}$, $\Delta\rho_{max}$ (e Å$^{-3}$) | -1.09, 0.72 | -0.43, 1.33 |
| $T_{min}$, $T_{max}$ | 0.154, 0.748 | 0.063, 0.542 |

**Table S2 | Atomic coordinates in the $CrBr_3$ crystal structure $R\overline{3}$ space group.**

| Atom | Site | Sym. | *x* | *y* | *z* | Occ. | U |
|---|---|---|---|---|---|---|---|
| Br | *18f* | 1 | 0.66601(12) | 0.67984(12) | 0.41117(4) | 1.000 | 0.018 |
| Cr1 | *6c* | 3. | 1.00000 | 1.00000 | 0.33362(12) | 0.878(4) | 0.011 |
| Cr2 | *3a* | -3. | 0.66667 | 1.33333 | 0.33333 | 0.244(8) | 0.024 |

**Table S3 | Atomic coordinates in the $CrBr_3$ crystal structure $P\overline{3}m1$ space group.**

| Atom | Site | Sym. | *x* | *y* | *z* | Occ. | U |
|---|---|---|---|---|---|---|---|
| Br | *2d* | 3m. | 0.33333 | 0.66667 | 0.2333(2) | 1.000 | 0.026 |
| Cr | *1a* | -3m. | 0.00000 | 0.00000 | 0.00000 | 0.667 | 0.018 |

**Table S4 | Summary of the pressure evolution of lattice parameters obtained in the high-pressure diffraction experiment on $CrBr_3$ single crystal.**

| Pressure (GPa) | *a (Å)* | *b (Å)* | *c (Å)* | *V (Å³)* | $R\overline{3}$ *conc.* | Space group |
|---|---|---|---|---|---|---|
| 0 (out DAC) | 6.3091(7) | 6.3091(7) | 18.342(2) | 632.27(12) | 0.60(5) | $R\overline{3}$ |
| 0.7 | 6.189(3) | 6.189(3) | 19.0(2) | 631(7) | 0.54(7) | $R\overline{3}$ |
| 3 | 6.1897(18) | 6.1897(18) | 17.70(4) | 587(1) | 0.50(9) | $R\overline{3}$ |
| 5.7 | 6.075(3) | 6.075(3) | 17.06(5) | 545(2) | 0.50(5) | $R\overline{3}$ |
| 6.4 | 6.026(2) | 6.026(2) | 16.84(4) | 530(1) | 0.42(8) | $R\overline{3}$ |
| 0 (out DAC) | 3.6439(17) | 3.6439(17) | 6.116(3) | 70.33(6) | 0.6(5) | $P\overline{3}m1$ |
| 0.7 | 3.6390(13) | 3.6390(13) | 6.10(4) | 70.0(4) | 0.54(7) | $P\overline{3}m1$ |
| 3 | 3.5822(6) | 3.5822(6) | 5.84(2) | 64.9(3) | 0.50(9) | $P\overline{3}m1$ |
| 5.7 | 3.5268(7) | 3.5268(7) | 5.58(3) | 60.1(3) | 0.50(5) | $P\overline{3}m1$ |
| 6.4 | 3.4975(5) | 3.4975(5) | 5.53(2) | 58.6(2) | 0.42(8) | $P\overline{3}m1$ |
| 8.4 | 3.494(2) | 3.494(2) | 5.41(7) | 57.1(7) | -0.1(7) | $P\overline{3}m1$ |
| 11.8 | 3.427(2) | 3.427(2) | 5.33(7) | 54.2(7) | 0.08(3) | $P\overline{3}m1$ |
| 16.2 | 3.344(3) | 3.344(3) | 5.36(9) | 52.0(8) | 0.01(1) | $P\overline{3}m1$ |
| 21.5 | 3.341(4) | 3.341(4) | 5.21(8) | 50.4(8) | -0.03(2) | $P\overline{3}m1$ |
| unloaded (outside DAC) | 3.6498(9) | 3.6498(9) | 6.1424(19) | 70.86(3) | 0.01(1) | $P\overline{3}m1$ |